\documentclass[12pt]{iopart}
\expandafter\let\csname equation*\endcsname\relax
\expandafter\let\csname endequation*\endcsname\relax
\usepackage[compress]{cite}
\usepackage{amsmath,amsfonts,amssymb,bbm, mathtools, mathrsfs}
\usepackage[utf8]{inputenc}
\usepackage{csquotes}
\usepackage{graphicx, color}
\usepackage{tikz,pgf,tikz-cd,float}
\usepackage{comment}
\usepackage{ulem}

\usepackage[colorlinks=true, % Enable colored links
            linkcolor=blue,  % Color for internal links (sections, equations, etc.)
            citecolor=cyan,   % Color for citation links
            urlcolor=teal, % Color for URLs
            pdfborder={0 0 0} % Remove border around links
           ]{hyperref}

\newcommand{\one}{\mathbbm 1}

\newcommand{\qmarks}[1]{``#1''}
\newcommand*\diff{\mathop{}\!\mathrm{d}}

\newcommand{\ket}[1]{|{#1}\rangle}
\newcommand{\bra}[1]{\langle{#1}|}
\newcommand{\braket}[2]{\langle{#1}|{#2}\rangle}

\begin{document}

\title[]{Relational Observables in Group Field Theory}

\author{Luca Marchetti}

\address{Department of Mathematics and Statistics, University of New Brunswick,\\
Fredericton, NB, Canada E3B 5A3\\
Okinawa Institute of Science and Technology Graduate University,\\
Onna, Okinawa 904 0495 Japan\\
Kavli Institute for the Physics and Mathematics of the Universe (WPI),\\ UTIAS, The University of Tokyo, Chiba 277-8583, Japan
}
\ead{luca.marchetti@oist.jp}

\author{Edward Wilson-Ewing}
\address{Department of Mathematics and Statistics, University of New Brunswick,\\
Fredericton, NB, Canada E3B 5A3 \\
Department of Physics, McGill University, \\
Montr\'eal, QC, H3A 2T8, Canada}
\ead{edward.wilson-ewing@unb.ca}
\vspace{10pt}
\begin{indented}
\item[]19 December 2024 %\today
\end{indented}

\begin{abstract}
We construct relational observables in group field theory (GFT) in terms of covariant positive operator-valued measures (POVMs), using techniques developed in the context of quantum reference frames. We focus on matter quantum reference frames; this can be generalized to other types of frames within the same POVM-based framework. The resulting family of relational observables provides a covariant framework to extract localized observables from GFT, which is typically defined in a perspective-neutral way. Then, we compare this formalism with previous proposals for relational observables in GFT. We find that our quantum reference frame-based relational observables overcome the intrinsic limitations of previous proposals while reproducing the same continuum limit results concerning expectation values of the number and volume operators on coherent states. Nonetheless, there can be important differences for more complex operators, as well as for other types of GFT states. Finally, we also use a specific class of POVMs to show how to project states and operators from the more general perspective-neutral GFT Fock space to a perspective-dependent one where a scalar matter field plays the role of a relational clock.
\end{abstract}

\newpage

\section{Introduction}
\label{sec:intro}

Background independence is a crucial property of gravitational theories, including general relativity. As a result of this property, physical states--—represented by specific field configurations—--remain invariant under the action of spacetime diffeomorphisms, which are gauge symmetries of the theory. Since, from the active perspective, diffeomorphisms effectively move points, the location of these points in the manifold has no intrinsic physical significance, implying that spacetime-localized quantities cannot be physical observables, as they are not gauge-invariant \cite{Torre:1993fq} (on the other hand global observables trivially are, for instance any integral over the whole spacetime of a covariant top form is gauge-invariant). 

The apparent tension between background independence and locality can be addressed by recognizing that gauge-invariant local observables can be constructed by describing dynamical quantities relative to one another \cite{Goeller:2022rsx}. These \textit{relational} observables are localized with respect to \textit{dynamical}, rather than external, reference frames. In the context of classical gravity, such observables were first introduced in \cite{bergmann1, bergmann2, bergmann3} and \cite{Rovelli_1991a, Rovelli_1991b, Gambini:2000ht, Rovelli:2001bz}, and later developed into more systematic and general constructions in \cite{Dittrich:2004cb, Dittrich:2005kc, Giesel:2007wi, Pons:2009cz, Tambornino:2011vg, Dapor:2013hca, Khavkine:2015fwa, Goeller:2022rsx}, showing in particular their covariance and their equivalence \cite{Carrozza:2021gju, Goeller:2022rsx, Carrozza:2022xut} to the \qmarks{dressed observables} \cite{Donnelly:2015hta, Donnelly:2016rvo, Giddings:2018umg, Giddings:2019hjc} frequently used in gauge theories. The use of relational observables has also been explored in cosmology \cite{Dittrich:2007jx, Giesel:2007wk, Husain:2011tm, Giesel:2017roz, Giesel:2018opa, Giesel:2018tcw, Giesel:2020bht, Giesel:2020xnb, Husain:2020uac}. At the quantum level, significant progress has been made in the implementation of the relational strategy within (constrained) quantum mechanical systems \cite{Hoehn:2019fsy, Hoehn:2020epv, Loveridge:2017pcv, Vanrietvelde:2018dit, Vanrietvelde:2018pgb, Hoehn:2021flk, delaHamette:2021oex, Carette:2023wpz, Hoehn:2023ehz, Busch:2016rui, Hausmann:2023jpn}, showing in particular the quantum covariance of physical properties under changes of quantum reference frames (QRFs) \cite{Hoehn:2019fsy, Hoehn:2020epv, Vanrietvelde:2018dit, Vanrietvelde:2018pgb, Hoehn:2021flk, delaHamette:2021oex, Giacomini:2017zju, Kabel:2024lzr}, as well as the dependence on QRFs of quantum correlations and (thermo)dynamical properties of subsystems \cite{Hoehn:2023ehz,Castro-Ruiz:2021vnq}. Moreover, recently, an analogous relational description has been implemented within parametrized quantum field theory by using (ideal) QRFs \cite{Hoehn:2023axh}. Despite this progress, a systematic implementation of the relational strategy and study of QRFs within the context of quantum field theory (QFT), even on a fixed background, is still lacking.

If constructing relational observables is already a difficult task in standard QFT, it becomes even more challenging in the context of quantum gravity (QG). This is because in many approaches to quantum gravity, spacetime notions become less well defined (and sometimes disappear entirely), making it extremely difficult to import insights from the classical, spacetime-based theory \cite{Oriti:2018}. Somewhat ironically, however, it is precisely for this reason that constructing relational observables becomes crucial: they often represent the only viable means of defining localization in QG. Consequently, relational observables have been a central focus of study across a wide range of QG theories, including string theory \cite{Gary:2006mw}, holography \cite{Heemskerk:2012np, Almheiri:2014lwa, Giddings:2018umg}, asymptotic safety \cite{Baldazzi:2021fye}, perturbative algebraic QG \cite{Brunetti:2013maa, Brunetti:2016hgw, Rejzner:2016yuy}, loop quantum gravity (LQG) \cite{Rovelli:1993bm, Thiemann:2004wk, Giesel:2007wn, Kaminski:2009qb, Domagala:2010bm, Husain:2011tk, Giesel:2012rb, Giesel:2016gxq, Giesel:2023now} and group field theory (GFT) \cite{Oriti:2016qtz, Gielen:2016dss, Gielen:2017eco, Gielen:2018xph, Gielen:2018fqv, Gerhardt:2018byq, Wilson-Ewing:2018mrp, Gielen:2019nyq, Gielen:2019kae, Gielen:2020fgi, Gielen:2023szb, Marchetti:2020umh, Marchetti:2020qsq, Gielen:2021vdd, Marchetti:2021gcv, Jercher:2021bie, Jercher:2023kfr, Jercher:2023nxa, Calcinari:2024pek}.

GFT exemplifies the duality of constructing relational observables in QG: as GFT is a QFT of spacetime quanta \cite{Freidel:2005qe, Oriti:2011jm, Krajewski:2011zzu}, in this approach spacetime notions are expected to emerge only in an appropriate regime \cite{Oriti:2024}; in particular, a continuum geometry can only emerge for highly-excited states \cite{Marchetti:2022nrf}. Therefore, the methods used to construct relational observables in GFT cannot simply replicate those from classical gravity. As a result, various approaches for constructing relational observables in GFT have been proposed in the literature. For example, in \cite{Oriti:2016qtz, Gielen:2016dss, Gielen:2017eco, Gielen:2018xph, Gielen:2018fqv, Gerhardt:2018byq}, relational observables are defined by localizing the (generally non-local) second-quantized operator constructed in the \qmarks{perspective-neutral} GFT Fock space with respect to a chosen sector of the GFT field domain, assumed to represent the physical frame. This construction was further developed in \cite{Wilson-Ewing:2018mrp, Gielen:2019nyq, Gielen:2019kae, Gielen:2020fgi, Gielen:2023szb, Calcinari:2024pek} by introducing a \qmarks{reduced} Fock space specifically designed to accommodate such observables. In contrast, \cite{Marchetti:2020umh, Marchetti:2020qsq, Marchetti:2021gcv, Jercher:2021bie, Jercher:2023kfr, Jercher:2023nxa, Gielen:2021vdd} take an effective approach, constructing relational observables by averaging perspective-neutral observables over collectively localized states (typically coherent peaked states).

While these methods appear to produce similar results in cosmological applications (compare, e.g., \cite{Oriti:2016qtz, Gielen:2019kae, Marchetti:2020umh}), they each exhibit certain limitations. For instance, the sharply localized observables introduced in \cite{Oriti:2016qtz, Gielen:2016dss, Gielen:2017eco, Gielen:2018xph, Gielen:2018fqv, Gerhardt:2018byq} face interpretational ambiguities, as the variables used for localization do not have an obvious connection with the operators representing the quantum frame in the (perspective-neutral) GFT Fock space \cite{Marchetti:2020umh}. This issue can be conceptually mitigated by focusing on a Fock space reduced by deparametrization \cite{Wilson-Ewing:2018mrp, Gielen:2019nyq, Gielen:2019kae, Gielen:2020fgi, Gielen:2023szb} (it has recently been shown that this deparametrized approach can be described covariantly by reparametrizing the theory \cite{Calcinari:2024pek}), although this approach relies on the assumption that the reduction to the frame perspective (i.e., the initial deparametrization) is always possible and does not significantly alter the structure of the Fock space. In many continuum systems, this assumption does not hold and it may be overly restrictive for GFT as well. The effective method of \cite{Marchetti:2020umh, Marchetti:2020qsq, Marchetti:2021gcv, Jercher:2021bie, Jercher:2023kfr, Jercher:2023nxa, Gielen:2021vdd} offers clearer physical interpretations of the relational observables but is limited to certain states and specific regimes.

In this work, we aim to overcome these limitations by adapting to the GFT formalism the QRF approach of building relational observables by conditioning system observables on the orientation of a reference frame. In this way, we can construct relational observables in GFT that: (i) allow for a clear physical interpretation as observables conditioned on quantum reference frame configurations; (ii) act on the perspective-neutral GFT Fock space; (iii) are defined independently of the quantum state; and (iv) coincide with previously defined relational observables when both are averaged over appropriate collective states, thereby ensuring that they can be successfully used to extract continuum physics, and in particular make contact with previous results in the cosmological setting. More specifically, to achieve this we introduce covariant positive operator-valued measures (POVMs) \cite{Busch:1997, Nielsen_Chuang_2010, Heinosaari_Ziman_2011} associated with minimally-coupled massless scalar field QRFs (here and throughout the paper, massless is understood to mean without any potential). Motivated by the standard QRF approach \cite{Hoehn:2019fsy, Hoehn:2020epv, Loveridge:2017pcv, Vanrietvelde:2018dit, Vanrietvelde:2018pgb, Hoehn:2021flk, delaHamette:2021oex, Carette:2023wpz, Hoehn:2023ehz, Busch:2016rui, Hausmann:2023jpn} (see also \cite{Glowacki:2024jhi}), we then use these POVMs to construct relational observables by (on-shell) conditioning perspective-neutral observables on frame orientations, a construction performed here for, to the best of our knowledge, the first time in full QG.

The outline of this work is as follows. First, in Sec.~\ref{sec:gfts}, we give an introduction to GFT, with a particular emphasis on the perspective-neutral Fock space in Sec.~\ref{sec:gftfock}. Then, in Sec.~\ref{sec:relobsforgft}, we construct relational observables for GFT that satisfy the four conditions outlined above. Specifically, in Sec.~\ref{sec:relobservables}, we review the definition and various approaches for constructing relational observables in both classical and quantum mechanical settings, while in Sec.~\ref{sec:povms}, we demonstrate how to construct POVMs associated with minimally-coupled massless scalar field QRFs in GFT, and in Sec.~\ref{sec:relationalobservablesgft} we utilize these POVMs to construct relational observables. In Sec.~\ref{sec:comparison}, we compare the newly defined relational observables with existing definitions in the literature: in Sec.~\ref{sec:coherentstates}, we compute the expectation values of the relational observables defined in Sec.~\ref{sec:relobsforgft} on coherent states and compare these with the corresponding averages of the sharply localized relational observables used in \cite{Oriti:2016qtz, Gielen:2016dss, Gielen:2017eco, Gielen:2018xph, Gielen:2018fqv, Gerhardt:2018byq}, while in Sec.~\ref{sec:effectiverelobs} we show how POVMs with intrinsic measurement uncertainty can produce relational observables whose averages on coherent states agree with the effective relational observables of \cite{Marchetti:2020umh, Marchetti:2020qsq, Marchetti:2021gcv, Jercher:2021bie, Jercher:2023kfr, Jercher:2023nxa}, and in Sec.~\ref{sec:deparametrized} we demonstrate how a specific class of POVMs can be used to construct the \qmarks{perspective-reduced} (or deparameterized) GFT Fock space of \cite{Wilson-Ewing:2018mrp, Gielen:2019nyq, Gielen:2019kae, Gielen:2020fgi, Gielen:2023szb}. We end with a discussion of the main results of this work and outline possible directions for future research in Sec.~\ref{sec:conclusions}.

\section{Group Field Theory}
\label{sec:gfts}

A GFT model is defined by the group domain $\mathcal{D}$ of the GFT field $\Phi:\mathcal{D}\to \mathbb{C}$ and the GFT action $S_{\text{GFT}}$. As in standard QFT on a spacetime, quantum amplitudes $\mathcal{A}_\Gamma$ can be described using graphs decorated with group-theoretic data associated with the domain $\mathcal{D}$. By appropriately choosing $\mathcal{D}$ and $S_{\text{GFT}}$ \cite{Freidel:2005qe}:
\begin{itemize}
    \item The Feynman graphs $\Gamma$ are dual to cellular complexes with arbitrary topology;
    \item The amplitudes $\mathcal{A}_\Gamma$ can be identified with simplicial gravity (in a Plebanski-like formulation, representing discrete gravitational quantities in terms of group-theoretic data) path integrals on a lattice dual to the graph $\Gamma$;
    \item If one uses representation data instead of describing $\mathcal{A}_\Gamma$ in terms of group-theoretic quantities, $\mathcal{A}_\Gamma$ take the form of spinfoam models.
\end{itemize}
In other words, GFT provides a way to define a generating functional (the GFT partition function) for simplicial gravity and spinfoam models that naturally includes a sum over topologies. Moreover, they do so using the language of QFT, which offers many powerful computational tools. However, this comes at the expense of a straightforward geometrical interpretation since, as we will explicitly see below, spacetime notions can only be emergent in GFT.

As a concrete example of a GFT model, consider a $4$-dimensional Lorentzian EPRL spinfoam model (originally defined for the vacuum case \cite{Engle:2007wy, Freidel:2007py}) minimally coupled with $m$ massless scalar fields \cite{Oriti:2016qtz, Li:2017uao, Marchetti:2021gcv}. This model, which we focus on in this paper (although the general strategy outlined below can, in principle, be applied to any GFT model), has been extensively studied in the past and used to extract continuum and cosmological physics \cite{Oriti:2016qtz, Oriti:2016ueo, Gielen:2019kae, Marchetti:2020qsq, Marchetti:2021gcv}. The domain $\mathcal{D}$ of $\Phi$ in this case is $\mathcal{D}=\mathrm{SU}(2)^4\times\mathbb{R}^m$, and excitations of the GFT field admit a clear geometric interpretation in terms of $3$-simplices (i.e., tetrahedra), once the following closure constraint is imposed \cite{Oriti:2011jm}:
\begin{equation}\label{eqn:closureconstraint}
    \Phi(\vec{g},\boldsymbol{\chi})=\Phi(\vec{g}\cdot_{\text{d}}h ,\boldsymbol{\chi})\,,\qquad \forall \, h\in\mathrm{SU}(2)\,,
\end{equation}
where $\vec{g}\cdot_\text{d} h\equiv \{g_1h,\dots,g_4h\}$ represents the diagonal right action of $h$ on $\vec{g}\equiv (g_1,\dots,g_4)\in\mathrm{SU}(2)^4$, and $\boldsymbol{\chi}=\{\chi^1,\dots,\chi^m\}\in\mathbb{R}^m$.

It is often useful to view the fundamental quanta of the theory as open spin-networks \cite{Oriti:2013aqa, Oriti:2014yla}, i.e., nodes from which emanate four links that are each decorated with the equivalence class of geometrical data $\{\vec{g}\}\equiv \{\vec{g}\cdot_{\text{d}}h,h\in\mathrm{SU}(2)\}$, and where the scalar fields $\boldsymbol{\chi}$ are located at the nodes. This interpretation can be made more explicit by working in the spin representation: expanding the field $\Phi(\vec{g},\boldsymbol{\chi})$ on a basis of functions of $L^2(G^4/G)$, where $G=\mathrm{SU}(2)$ 
\begin{align}\label{eqn:grouptospin}
    \Phi(\vec{g},\boldsymbol{\chi})&=\sum_\iota\sum_{\vec{j}}\sum_{\vec{m},\vec{n}}\Phi^{j_1,\dots,j_4;\iota}_{m_1,\dots,m_4}(\boldsymbol{\chi})\left[\prod_{i=1}^4\sqrt{d(j_i)}D^{j_i}_{m_in_i}(g_i)\right]\mathcal{I}^{j_1,\dots,j_4;\iota}_{n_1,\dots,n_4}\nonumber\\
    &\equiv \sum_{\vec{\kappa}}\Phi_{\vec{\kappa}}(\boldsymbol{\chi}) \, \psi_{\vec{\kappa}}(\vec{g})\,,
\end{align}
where $D^{j}_{mn}(g)$ are Wigner rotation matrices, $d(j)=2j+1$, and $\mathcal{I}^{j_1,\dots,j_4;\iota}_{n_1,\dots,n_4}$ is an $\mathrm{SU}(2)$ intertwiner that appears due to the right-invariance imposed by equation \eqref{eqn:closureconstraint}. It is precisely due to the choice of $\mathrm{SU}(2)^4$ as the gravitational part of the group domain together with equation \eqref{eqn:closureconstraint} that the field modes are decorated with spin network data $\vec{\kappa}\equiv \{\vec{j},\vec{m},\iota\}$, where $\vec{j}$ and $\vec{m}$ are respectively spin and angular momentum labels, while $\iota$ is an intertwiner quantum number.

A GFT action $S_{\text{GFT}}$ is generally composed of a kinetic term $K$, 
whose inverse is the propagator of the theory, and an interaction term $U+\bar{U}$, which determines how the tetrahedra are glued together to construct the desired 4-cell that serves as the basic building block for the discrete four-dimensional spacetime. The interaction term is typically characterized by a \qmarks{simplicial} form, i.e., it involves 5 powers of the GFT field $\Phi$, with the geometric data convoluted according to the combinatorial pattern dictated by the gluing of 5 tetrahedra to form a 4-simplex. Then,
\begin{equation}
    S_{\text{GFT}}=K+U+\bar{U}\,,
\end{equation}
where
\begin{subequations}\label{eqn:generickineticandint}
    \begin{align}
        K& = \int \diff\vec{g} \diff\vec{g}^{{}\,\prime} \int \diff\boldsymbol{\chi} \diff\boldsymbol{\chi}' \, \bar{\Phi}(\vec{g},\boldsymbol{\chi}) \, \mathcal{K}(\vec{g},\vec{g}^{{}\,\prime};\boldsymbol{\chi},\boldsymbol{\chi}') \, \Phi(\vec{g}^{{}\,\prime},\boldsymbol{\chi}')\,, \\
        U& = \int\left[\prod_{i=1}^5\diff\vec{g}_i \, \diff \boldsymbol{\chi}_i\right] \mathcal{U}(\vec{g}_1,\dots,\vec{g}_5;\boldsymbol{\chi}_1,\dots,\boldsymbol{\chi}_5) \left[\prod_{i=1}^5\Phi(\vec{g}_i,\boldsymbol{\chi}_i)\right].
    \end{align}
\end{subequations}
The precise form of $S_{\text{GFT}}$, encoded in the kernels $\mathcal{K}$ and $\mathcal{U}$, depends on the specific spinfoam (or simplicial gravity) model coupled with $m$ massless scalar fields that is considered. In particular, in the EPRL case, $\mathcal{K}$ and $\mathcal{U}$ encode the details of the imposition of the simplicity constraint, describing the embedding of the $\mathrm{SU}(2)$ part of the GFT field domain within the Lorentz group $\mathrm{SL}(2,\mathbb{C})$ \cite{Gielen:2013naa}. Additionally, $\mathcal{K}$ and $\mathcal{U}$ may be constrained by symmetries satisfied by the classical theory that are expected to be preserved during the quantization process. Examples of these symmetries include the reflection, shift, and rotation symmetries of minimally-coupled massless scalars: $\boldsymbol{\chi}\to-\boldsymbol{\chi}$, $\boldsymbol{\chi}\to\boldsymbol{\chi}+\boldsymbol{c}$, and $\boldsymbol{\chi}\to M\boldsymbol{\chi}$, with $M\in\mathrm{SO}(m)$ and $\boldsymbol{c}\in\mathbb{R}^m$ \cite{Oriti:2016qtz, Marchetti:2021gcv}. With these symmetries,
\begin{subequations}\label{eqn:reducedkineticandint}
    \begin{align}
        K &= \int \diff\vec{g} \diff\vec{g}^{{}\,\prime} \int \diff\boldsymbol{\chi} \diff\boldsymbol{\chi}' \, \bar{\Phi}(\vec{g},\boldsymbol{\chi}) \mathcal{K}(\vec{g},\vec{g}^{{}\,\prime};\vert\boldsymbol{\chi}-\boldsymbol{\chi}'\vert) \Phi(\vec{g}^{{}\,\prime},\boldsymbol{\chi}')\,,\\
        U &= \int\diff\boldsymbol{\chi} \left[\prod_{i=1}^5\diff\vec{g}_i\right] \mathcal{U}(\vec{g}_1,\dots,\vec{g}_5) \left[\prod_{i=1}^5\Phi(\vec{g}_i,\boldsymbol{\chi})\right].
    \end{align}
\end{subequations}
Despite the substantial simplifications introduced by the imposition of these symmetries, the resulting GFT remains quite complex, and calculations are difficult. For this reason, it is necessary to introduce approximations to extract physics; for example, in cosmological contexts one typically employs mean-field techniques and restricts the analysis to a mesoscopic regime where interactions are negligible \cite{Oriti:2016qtz, Oriti:2016ueo, Marchetti:2020qsq, Oriti:2024}.

Further details concerning the dynamical structure of GFT will not be needed, as in the remainder of our work we do not restrict ourselves to a specific GFT action.

\subsection{The perspective neutral GFT Fock space}
\label{sec:gftfock}

Although historically first introduced in a path-integral formulation (due to the connection with spinfoam models and simplicial gravity explained above) \cite{Reisenberger:2000zc}, GFT can also naturally be formulated in a canonical language \cite{Oriti:2013aqa, Gielen:2024sxs}; in particular, the path-integral and the Fock representation of GFT outlined below have been recently shown to be equivalent for a GFT formulation of the Husain-Kucha\v{r} theory \cite{Marchetti:2024tjq}, further confirming that the GFT Fock representation is naturally associated with a free theory—a result consistent with standard QFTs (the proof of this equivalence is possible since the GFT is non-interacting, see \cite{Kegeles} for a possible construction in the interacting case). To achieve this reformulation, creation $\hat{\varphi}^\dagger(\vec{g},\boldsymbol{\chi})$ and annihilation $\hat{\varphi}(\vec{g},\boldsymbol{\chi})$ operators are introduced that satisfy the following commutation relations:
\begin{subequations}\label{eqn:commutationrelationsg}
    \begin{align}
        [\hat{\varphi}(\vec{g},\boldsymbol{\chi}),\hat{\varphi}^\dagger(\vec{g}^{{}\,\prime},\boldsymbol{\chi}')]&=\one_G(\vec{g},\vec{g}^{{}\,\prime})\delta(\boldsymbol{\chi}-\boldsymbol{\chi}')\,,\\
        [\hat{\varphi}(\vec{g},\boldsymbol{\chi}),\hat{\varphi}(\vec{g}^{{}\,\prime},\boldsymbol{\chi}')]&=[\hat{\varphi}^\dagger(\vec{g},\boldsymbol{\chi}),\hat{\varphi}^\dagger(\vec{g}^{{}\,\prime},\boldsymbol{\chi}')]=0\,,
    \end{align}
\end{subequations}
where $\one_G$ is the Dirac delta distribution in the space $G^4/G$, with $G=\mathrm{SU}(2)$. Upon decomposition in representation labels, these commutation relations become
\begin{subequations}\label{eqn:commutationrelationsk}
    \begin{align}
        [\hat{\varphi}_{\vec{\kappa}}(\boldsymbol{\chi}),\hat{\varphi}^\dagger_{\vec{\kappa}'}(\boldsymbol{\chi}')]&=\delta_{\vec{\kappa},\vec{\kappa}'}\delta(\boldsymbol{\chi}-\boldsymbol{\chi}')\,,\\
        [\hat{\varphi}_{\vec{\kappa}}(\boldsymbol{\chi}),\hat{\varphi}_{\vec{\kappa}'}(\boldsymbol{\chi}')]&=[\hat{\varphi}^\dagger_{\vec{\kappa}}(\boldsymbol{\chi}),\hat{\varphi}_{\vec{\kappa}'}^\dagger(\boldsymbol{\chi}')]=0\,.
    \end{align}
\end{subequations}
There exist two proposals to relate these creation and annihilation operators to GFT field operators $\hat\Phi_{\vec{\kappa}}(\boldsymbol{\chi})$. In the case that $\Phi$ is a complex field, it is possible to directly identify $\hat\Phi_{\vec{\kappa}}(\boldsymbol{\chi}) = \hat \varphi_{\vec{\kappa}}(\boldsymbol{\chi})$ and $\hat \Phi^\dagger_{\vec{\kappa}}(\boldsymbol{\chi}) = \hat\varphi^\dagger_{\vec{\kappa}}(\boldsymbol{\chi})$ \cite{Oriti:2013aqa, Gielen:2024sxs}, while if $\Phi$ is real, then it can be decomposed into creation and annihilation operators as $\hat \Phi_{\vec{\kappa}}(\boldsymbol{\chi}) = (1 /\sqrt{2}) [\hat{\varphi}_{\vec{\kappa}}(\boldsymbol{\chi}) + \hat{\varphi}^\dagger_{\vec{\kappa}}(\boldsymbol{\chi})]$ \cite{Marchetti:2024tjq}. The choice of this identification does not affect the formalism for relational observables that we introduce in this paper, so we will remain agnostic on this point and simply focus on the GFT creation and annihilation operators in the following.

As in standard QFT, a Fock space $\mathcal{F}$ can be constructed by the repeated action of the creation operator $\hat{\varphi}^\dagger_{\vec{\kappa}}(\boldsymbol{\chi})$ on the Fock vacuum $\ket{0}$. We refer to $\mathcal{F}$ as a \qmarks{perspective neutral} Fock space, since equations \eqref{eqn:commutationrelationsg} and \eqref{eqn:commutationrelationsk} do not admit a preferred choice of a physical frame, cfr.\ equation \eqref{eqn:sametimecommutationrelations}. The term \qmarks{perspective neutral} is used here to highlight that the physical interpretation of $\mathcal{F}$ is analogous to that of the perspective neutral space used in the context of QRFs \cite{Vanrietvelde:2018pgb, Vanrietvelde:2018dit, Hohn:2018toe, Hoehn:2019fsy, Hoehn:2020epv, AliAhmad:2021adn, Hoehn:2021flk, Giacomini:2021gei, delaHamette:2021oex}, even though the technical construction differs for two important reasons.

A first reason regards the imposition of gauge-invariance. In the language of QRFs, the relevant symmetry group for quantum gravity would be (small) diffeomorphisms. In the GFT quantization of gravity, the continuum manifold associated with this symmetry is eliminated entirely, and the theory is instead defined on a domain given by the possible values (images) of the fields (both matter and gravitational). As a result, diffeomorphism invariance cannot be imposed directly as a symmetry of the theory; rather, the consequences of this classical symmetry are encoded in the intrinsically background-independent nature of the GFT quantization. In this sense, the GFT Fock space is not obtained through the explicit imposition of a gauge constraint, as is typically the case in QRF frameworks. However, analogously to the QRF case, \textit{this space encodes and links all possible internal frame perspectives}. Indeed, all fields that may be used as internal frames are included in the GFT field domain and are treated on an equal footing in equations (6) and (7). By contrast, equation (68) involves “same-time” commutation relations that induce a Fock space structure adapted to the perspective of a specific frame—more precisely, a chosen clock. This is the reason why we refer to the former as “perspective-neutral” and the latter as “perspective-reduced.” The perspective-neutral nature of $\mathcal{F}$ is further confirmed by the ability to derive perspective-reduced structures from it, as shown explicitly in Sec.\ \ref{sec:deparametrized}.

A second point concerns the Hilbert space structure of the GFT Fock space. While a distinction between “frame” and “system” degrees of freedom can be made at the level of the GFT field domain—where classical fields can be clearly separated due to the factorized structure of the domain—this factorization does not persist at the quantum level. In the full quantum theory, both the frame and system degrees of freedom are expected to emerge collectively from the behavior of the GFT field. Note that this lack of factorization is, in fact, to be expected \textit{after} the imposition of the quantum constraint. Both these points will be relevant in the construction of relational observables in Sec.~\ref{sec:relationalobservablesgft}.

As an example of states in $\mathcal{F}$ one can consider the one-particle state
\begin{equation}
    \ket{\vec{\kappa};\boldsymbol{\chi}}\equiv\hat{\varphi}^\dagger_{\vec{\kappa}}(\boldsymbol{\chi})\ket{0},
\end{equation}
which represents an open spin-network vertex with spin labels $\vec\kappa = (\iota, j_i, m_i)$ and scalar field data $\boldsymbol{\chi}=(\chi^1,\dots,\chi^m)$. Generic $n$-particle states correspond to a collection of $n$ indistinguishable spin-network vertices. To make the connection with canonical LQG, an important subspace of the Fock space is the space where the vertices are connected via entangled links (specifically, the $m$ labels for each pair of two entangled/connected links form a singlet state \cite{Oriti:2013aqa, Colafranceschi:2020ern}); these states can be identified with a graph, and therefore with a spin-network when taking into account the remaining $j$ and $\iota$ $SU(2)$ labels. (Note however that there is a difference between this subspace of the GFT Fock space and the Hilbert space of canonical LQG, as in canonical LQG the spin-networks are embedded in a three-dimensional manifold, in contrast to the abstract graphs that arise in GFT.) As a result, many states $\ket{\psi}\in\mathcal{F}$ do not correspond to a connected simplicial lattice---it is only states with the entanglement necessary to be fully connected that satisfy this property \cite{Oriti:2013aqa} (see also \cite{Oriti:2015qva} for a concrete construction of some GFT states that can be associated with corresponding LQG kinematical states). Therefore, from the perspective of GFT, the recovery of continuum quantities is a highly non-trivial process, possibly associated with emergent behavior.

Emergent, large-scale behavior can be effectively studied in GFT by constructing macroscopic quantities and collective states. The former are usually described in terms of second-quantized operators, which in general take the form
\begin{align}\label{eqn:nmbodyobs}
    \hat{O}^{(n,m)} &= \int \left[ \prod_{i=1}^{n} \diff\vec{g}_i \diff\boldsymbol{\chi}_i \, \hat{\varphi}^\dagger(\vec{g}_i,\boldsymbol{\chi}_i) \right] \left[ \prod_{j=1}^m \diff\vec{g}_j^{{}\,\prime} \diff\boldsymbol{\chi}_j' \, \hat{\varphi}(\vec{g}'_j,\boldsymbol{\chi}'_j) \right] O^{(n,m)},
\end{align}
where the matrix elements $O^{(n,m)}$ in principle can depend on all $\vec g_i, \boldsymbol{\chi}_i, \vec{g}_j^{{}\,\prime}, \boldsymbol{\chi}_j'$. 

Particularly relevant examples, especially for cosmological applications, are the total number and volume operators, as well as the operators corresponding to the (extensive) scalar fields and their momenta, given respectively by
\begin{subequations}
    \begin{align}\label{eqn:numberoperator}
        \hat{N}&=\int\diff\vec{g}\diff\boldsymbol{\chi} \,\, \hat{\varphi}^\dagger(\vec{g},\boldsymbol{\chi})\hat{\varphi}(\vec{g},\boldsymbol{\chi})\,,\\
        \hat{V}&=\int\diff\vec{g}\diff\vec{g}^{{}\,\prime}\diff\boldsymbol{\chi} \,\, \hat{\varphi}^\dagger(\vec{g},\boldsymbol{\chi})\mathcal{V}(\vec{g},\vec{g}^{{}\,\prime})\hat{\varphi}(\vec{g}^{{}\,\prime},\boldsymbol{\chi})\,, \\
        \hat{X}^a&=   \int\diff\vec{g}\diff\boldsymbol{\chi}\,\chi^a\hat{\varphi}^\dagger(\vec{g},\boldsymbol{\chi})\hat{\varphi}(\vec{g},\boldsymbol{\chi})\,,\\
        \hat{\Pi}_a&=-i   \int\diff\vec{g}\diff\boldsymbol{\chi}\,\hat{\varphi}^\dagger(\vec{g},\boldsymbol{\chi})\partial_{\chi^a}\hat{\varphi}(\vec{g},\boldsymbol{\chi})\,,
    \end{align}
\end{subequations}
where the kernel $\mathcal{V}(\vec{g},\vec{g}^{{}\,\prime})$ of the volume operator is obtained from the matrix elements of the LQG volume operator on spin-network states \cite{Oriti:2016qtz}.

As can immediately be seen from the above expressions, these operators are not localized in any sense, and do not immediately admit a relational interpretation. Nonetheless, relational observables can be constructed from these operators, as shown below in Sec.\ \ref{sec:relobservables}.

Finally, collective (non-perturbative) states also play a crucial role in extracting continuum physics from GFT. Particular attention has been devoted to coherent states
\begin{equation}\label{eqn:coherentstates}
    \ket{\sigma} = \mathcal{N}_\sigma \exp\left[ \sum_{\vec{\kappa}} \int \diff\boldsymbol{\chi} \, \sigma_{\vec{\kappa}}(\boldsymbol{\chi}) \hat{\varphi}^\dagger_{\vec{\kappa}}(\boldsymbol{\chi}) \right] |0\rangle \, ,
    \quad \vert\mathcal{N}_\sigma\vert^2 = \exp \left[ -\sum_{\vec{\kappa}} \int \diff\boldsymbol{\chi}\, \vert \sigma_{\vec{\kappa}}(\boldsymbol{\chi}) \vert^2 \right],
\end{equation}
these are eigenstates of the annihilation operator $\hat{\varphi}_{\vec{\kappa}(\chi)}$, i.e., $\hat{\varphi}_{\vec{\kappa}(\chi)}\ket{\sigma}=\sigma_{\kappa}(\chi)\ket{\sigma}$.

Coherent states have been used to construct cosmological states: by imposing homogeneity and isotropy on $\sigma$, one can recover cosmological physics in an appropriate limit \cite{Oriti:2016qtz, Oriti:2016ueo, Gielen:2016dss, Gielen:2019kae, Gielen:2020fgi, Marchetti:2020umh, Jercher:2021bie}; further, an analogous matching has been achieved even for slightly inhomogeneous cosmologies by enriching the states \eqref{eqn:coherentstates} with quantum correlations \cite{Jercher:2023kfr, Jercher:2023nxa}.

Given their crucial role in the study of continuum physics in GFT, Sec.~\ref{sec:coherentstates} will be devoted to the study of relational observables evaluated on coherent states.

\section{Relational observables in Group Field Theory}
\label{sec:relobsforgft}

\subsection{Relational observables}
\label{sec:relobservables}

As relational observables are essential for describing the physics of background-independent theories, they have been the subject of extensive investigation in recent years and are now relatively well understood, particularly at the classical level. Relational observables essentially encode information about some dynamical fields relative to other dynamical fields (often called frame fields).

To be more explicit, following \cite{Goeller:2022rsx}, let us denote the spacetime manifold by $\mathcal{M}$, and the kinematical field configurations of the theory by $\phi$. Then, a frame field $R^{-1}[\phi]:\mathcal{N}[\phi]\to \mathcal{O}$ is a map from a certain region $\mathcal{N}[\phi]\in\mathcal{M}$ to a region $\mathcal{O}$ of the frame orientation space that, under a small diffeomorphism $f$ satisfies
\begin{equation}
    R[\phi]^{-1}\mapsto R[f_*\phi]^{-1}=R[\phi]^{-1}\circ f^{-1}\,,
\end{equation}
where $f_*$ represents the push-forward of $f$. In other words, the frame field provides a map between a kinematical (or coordinate) notion of locality and a physical one \cite{Goeller:2022rsx}. As a simple example, if a scalar field $\chi$ is used as a frame field, $R^{-1}$ maps any slice $\chi=\chi_0$ in the manifold to the corresponding subset in configuration space.

Given a local covariant observable $O$, the relational observable constructed by pushing forward $O$ through the frame $R[\phi]^{-1}$, namely
\begin{equation}
    O_R[\phi]\equiv (R[\phi]^{-1}_*)O[\phi]\,,
\end{equation}
is invariant under small diffeomorphisms \cite{Goeller:2022rsx}. This construction can be shown to be equivalent (under some assumptions) to other representations of relational observables \cite{Goeller:2022rsx}, including the \qmarks{single-integral} ones \cite{Gary:2006mw}
\begin{equation}\label{eqn:singleintegral}
    O_Z[\phi](\chi_0)=\int_{\mathcal{M}} \diff^Dx \,\, \delta^{(D)}(\chi_0^a-\chi^a) \left \vert \det \frac{\partial \chi^a}{\partial x^\mu} \right \vert O(x)\,,
\end{equation}
where $\chi^a$, $a=0,\dots, D-1$, are $D$ scalar fields, which are collectively used as the physical frame field $R^{-1}=\chi\equiv (\chi^0,\dots, \chi^{D-1})$, and also the canonical ones, written in series form
\begin{align}\label{eqn:canonical}
    O_{Z}[\phi](Z_0)&=\sum_{n=0}^\infty\int\diff^{D-1}\mathbf{x}_1\cdots\diff^{D-1}\mathbf{x}_n\, G^{a_1}(\mathbf{x}_1)[\chi;\chi_0]\cdots G^{a_n}(\mathbf{x}_n)[\chi;\chi_0]\nonumber\\
    &\quad \qquad \times\left\{\dots\left\{O(\mathbf{x}_1),\tilde{C}_{a_1}(\mathbf{x}_1)\right\}\dots, \tilde{C}_{a_n}(\mathbf{x}_n)\right\},
\end{align}
where $G^{\mu}(\mathbf{x})=\chi^a_0-\chi^a$ are gauge-fixing conditions depending on the frame field $R^{-1}=\chi$, $\tilde{C}_a$ are (weakly abelianized) constraints generating the gauge symmetries of the theory, $\mathbf{x}$ are coordinates on the chosen Cauchy slice $\Sigma$, and $\{\cdot,\cdot\}$ are kinematical Poisson brackets \cite{Dittrich:2004cb,Dittrich:2005kc}.

At the quantum level, significant progress has been made for mechanical systems, for any unimodular gauge symmetry group $G$ \cite{delaHamette:2021oex}. The relevant Hilbert space, as described from the perspective of an external frame, is assumed to be of the form $\mathcal{H}_{\text{kin}}=\mathcal{H}_R\otimes\mathcal{H}_S$, thus factorizing into a quantum reference frame Hilbert space $\mathcal{H}_R$ and a system one, $\mathcal{H}_S$. Moreover, one assumes that $\mathcal{H}_{\text{kin}}$ carries a unitary tensor product representation of the symmetry group $G$: $\mathcal{H}_{\text{kin}}\ni\ket{\psi}_{\text{kin}}\mapsto U_{RS}(h)\ket{\psi}_{\text{kin}}=U_R(g)\otimes U_S(g)\ket{\psi}_{\text{kin}}$.

In this context, the orientations of the quantum reference frame can be represented by $G$-coherent states $\ket{\phi(g)}_R$, generated by the action of the group on a \qmarks{seed} state $\ket{\phi(e)}\in\mathcal{H}_R: \ket{\phi(g)}=U_R(g)\ket{\phi(e)}$ \cite{Perelomov:1986}. Assuming that $G$ acts transitively and freely on the coherent state system, there is an invertible equivariant map $e:G\to \mathcal{H}_R$, so that the above frame orientations can completely parametrize the gauge orbits in the Hilbert space or observable algebra \cite{delaHamette:2021oex}. Denoting the set of linear operators on the Hilbert space $\mathcal{H}$ by $\mathcal{L}(\mathcal{H})$, then for any operator $\hat{O}^{(S)}\in\mathcal{L}(\mathcal{H}_S)$, the corresponding relational observable is \cite{delaHamette:2021oex} (and see also \cite{Busch:2016rui, Loveridge:2017pcv, Carette:2023wpz, Fewster:2024pur})
\begin{equation}\label{eqn:operativerelational}
    \hat{O}^{(S)}_{R}=\mathcal{G}\left(\ket{\phi(g)}_R\bra{\phi(g)}_R\otimes \hat{O}^{(S)}\right),
\end{equation}
where the $G$-twirl, defined for all operators $A \in \mathcal{L}(\mathcal{H}_{\text{kin}})$ as
\begin{equation}\label{eqn:gtwirl}
    \mathcal{G}(A)=\int_G \diff h \,\, U_{RS}(h) \, A \,\, U_{RS}(h)^\dagger\,,
\end{equation}
is an incoherent group-averaging. In other words, the relational observable $O_R^{(S)}$ is constructed by first tensoring the kinematical system observable $O^{(S)}$ with the projector $\ket{\phi(g)}_R\bra{\phi(g)}_R$ onto the orientation state of the quantum reference frame (called frame orientation conditioning), and then performing the $G$-twirl that guarantees that the relational observable $O_S^{(S)}$ is gauge-invariant. Note that the integral in \eqref{eqn:gtwirl} generally does not converge for non-compact groups; see \cite{delaHamette:2021oex} for a more detailed discussion. However, as we will see in Sec.~\ref{sec:relationalobservablesgft}, the construction of relational observables in GFT does not require such group averaging, thereby avoiding such convergence issues.

\subsection{POVMs in QFT and GFT}
\label{sec:povms}

In quantum theories, positive operator-valued measures (POVMs) \cite{Busch:1997, Nielsen_Chuang_2010, Heinosaari_Ziman_2011} are naturally associated with measurement processes.  Intuitively, POVMs provide a set of operators $\hat E_i$ adapted to a particular measurement that can be used to calculate the probability of each possible outcome of that measurement on any state $|\psi\rangle$, with the probability of the $i$-th possible outcome given by $p_i = \langle \psi | \hat E_i | \psi \rangle$.

A more precise definition, and one that can also be used in quantum field theory, is the following:
Let $(\Omega, \Sigma)$ be a measurable space, with $\Omega$ a non-empty set and $\Sigma$ a $\sigma$-algebra of subsets of $\Omega$. Given an operator algebra $\mathcal{L}(\mathcal{H})$ on a Hilbert space $\mathcal{H}$, a positive operator valued measure (POVM) is a map $\hat{E}:\Sigma\to\mathcal{L}(\mathcal{H})$ that satisfies \cite{Busch:1997}
\begin{enumerate}
    \item $\hat{E}(X) \ge \hat{O}$ for all $X\in\Sigma$, where $\hat{O}$ is the null operator,
    \item For any countable collection of disjoint sets $X_i\subset\Sigma$, $\hat{E}(\cup_iX_i)=\sum_i\hat{E}(X_i)$, (with convergence in the weak operator topology required in the case of an infinite series).
\end{enumerate}
Operators $\hat{E}(X)$ in the range of POVMs are called \textit{effects} \cite{Busch:1997, Carette:2023wpz}. If the POVM also satisfies
\begin{enumerate}\setcounter{enumi}{2}
    \item $\hat{E}(\Omega)=\hat{\one}$, where $\hat{\one}$ is the identity operator,
\end{enumerate}
then the POVM is said to be \textit{normalized}, and is understood to cover all possible measurement outcomes. Together, these three properties ensure (the expectation values of) the POVMs can be interpreted as probabilities; note that $\Omega$ is to be understood as the space containing all possible outcomes of the measurement of interest, performed on any state $|\psi\rangle$ in the Hilbert space $\mathcal{H}$.

Furthermore, let $G$ be a locally compact second countable topological group, $.: G\times \Omega \to\Omega$ a continuous transitive action (so that $\Omega$ is a homogeneous $G$-space) and $U: G \mapsto \mathcal{L}(H)$ a strongly continuous projective unitary representation. Then, a POVM $\hat{E}:\Sigma\to \mathcal{L}(\mathcal{H})$ is called $G$-\textit{covariant} if it satisfies \cite{Holevo1, Holevo2, Holevo:1982book, Werner1986, Werner1987, Busch:1997, Holevo:2001book} (see also \cite{Hoehn:2019fsy, delaHamette:2021oex, Carette:2023wpz} for applications in the context of QRFs):
\begin{enumerate}\setcounter{enumi}{3}
    \item $\hat{E}(g.X)=\hat{U}(g)\hat{E}(X)\hat{U}^\dagger(g)$, for all $X\in \Sigma$ and for all $g\in G$.
\end{enumerate}

As a concrete example of the above definition, the orientation states $\ket{\phi(g)}_R$ used in \eqref{eqn:operativerelational} provide a resolution of the identity on $\mathcal{H}_R$, and thus give rise to the following POVM
\begin{equation}
    \hat{E}_\phi(Y)=\int_Y\diff g\,\ket{\phi(g)}\bra{\phi(g)}_R,
\end{equation}
for all $Y\in\mathcal{B}(G)$, where $\mathcal{B}(G)$ represents the Borel sets on the group $G$. By construction, this POVM is $G$-covariant, meaning that
\begin{equation}\label{eqn:covariance}
    \hat{E}_\phi(h.Y)=U_R(h)\hat{E}_\phi(Y)U_R^\dagger(h)\,,
\end{equation}
where, as before, $h.Y$ denotes the action of $h$ on $Y$. This shows that the introduction of POVMs provides a notion of probability distribution for the frame orientation with respect to which one can condition kinematical observables in order to obtain relational (i.e., physical) observables. This operative (and POVM-based) approach has been adopted to take some first steps towards constructing relational observables in quantum field theory (QFT) \cite{Glowacki:2024jhi}.

Measurement processes, as captured by POVMs, are typically studied by dynamically coupling the quantum fields of interest to appropriate ``probe" systems, which can be classical, quantum mechanical, or even quantum field theoretical \cite{Anastopoulos:2021nee, Fewster:2023cfq}. The probes are then investigated to determine how they provide information about the quantum fields being measured. Many models for QFT measurements use Unruh-DeWitt detectors \cite{Unruh:1976, DeWitt:1980hx, Crispino:2007eb}, where the quantum field is coupled to a point-like detector \cite{Perche:2022dzy, Hu:2012jr, Polo-Gomez:2021irs}. Despite the limitation of considering the detector as an intrinsically quantum mechanical system, these models allow simpler and more explicit computations, including concrete constructions of the POVM associated with the measurement process \cite{Polo-Gomez:2021irs, Terno:2001ue}.

Within the algebraic approach to QFT, there has been developed a framework for describing measurement processes where the probe is itself a quantum field. However, within this framework, it is significantly more challenging to perform explicit computations, extract information from the probe field \cite{Grimmer:2021qib}, and analyze concrete scenarios \cite{Anastopoulos:2021nee}. Indeed, an appropriate description of low-energy measurements requires going beyond free field theory, and the treatment of bound states in QFT remains an open problem.

Event POVMs in Minkowski spacetime have been investigated in QFT to achieve a consistent quantization of spacetime coordinates \cite{Toller:1997pc, Toller:1998wf, Toller:1998vk, Giannitrapani:1998pm, Mazzucchi:2001}, 
(time quantization has also been investigated in quantum mechanics, see, e.g., \cite{Holevo:2011, Busch:1997, Busch:1994, Giannitrapani:1996sq}). For this definition to be consistent, event POVMs must transform covariantly under the Poincaré group, see condition (iv) in the definition of POVMs given above. It is also important to note that, unlike standard POVMs, event POVMs in QFT cannot be normalized, as the vacuum state cannot define any event \cite{Toller:1998wf}.  

Similar challenges, possibly exacerbated by the less transparent physical interpretation of the GFT field, would arise when attempting to describe a measurement process of the GFT field. However, the context is different in GFT: Eq.~\eqref{eqn:operativerelational} suggests that to construct relational observables in GFT, the focus should be on probability measures associated with the frame fields rather than the GFT field itself, as shall be discussed in more detail in Sec.\ \ref{sec:relationalobservablesgft}. Frame fields, whether constructed from geometric quantities or matter fields, become part of the GFT field domain following the GFT quantization process. Thus, constructing frame POVMs in GFT is technically analogous to constructing indirect event POVMs in spacetime \cite{Toller:1997pc, Toller:1998wf}. 

Let us illustrate how frame POVMs can be constructed in the Fock representation of GFT by defining POVMs for the measurement outcomes of a single massless scalar field (that can be interpreted as, for example, a relational clock); in this case, the space of possible measurement outcomes $\Omega$ (see the definition of POVMs above) is simply $\mathbb{R}$ corresponding to the possible values that can be attained by the scalar field $\chi$.

Different definitions for the effect density are possible, so long as they satisfy the first two properties (i) and (ii) for POVMs; however, a natural choice that has nice properties is the following effect density (for another possible choice, see Sec.~\ref{sec:deparametrized}):
\begin{align}\label{eqn:gftpovmreal}
    \hat{E}_\chi(\diff\chi)&=\diff\chi\sum_{n=1}^\infty\frac{1}{n!} \sum_{\vec{\kappa}_1,\dots\vec{\kappa}_n} \! \int \! \left[ \prod_{i=1}^n\diff\chi_i \right]\frac{\sum_{i=1}^n\delta(\chi_i-\chi)}{n}\left[\prod_{i=1}^n\hat{\varphi}_{\vec{\kappa}_i}^\dagger(\chi_i)\right]\!\ket{0}\bra{0}\!\left[\prod_{i=1}^n\hat{\varphi}_{\vec{\kappa}_i}(\chi_i)\right]\nonumber\\
    &=\diff\chi\sum_{n=1}^\infty\sum_{\vec{\kappa}_1,\dots\vec{\kappa}_n}\int\left[\prod_{i=1}^n\diff\chi_i\right]\frac{\sum_{i=1}^n\delta(\chi_i-\chi)}{n}\hat{F}^{(n)}_{\vec{\kappa}_1,\dots,\vec{\kappa}_n}(\chi_1,\dots,\chi_n) \,,
\end{align}
where we have used \eqref{eqn:grouptospin} and introduced the quantities
\begin{equation}
    \hat{F}^{(n)}_{\vec{\kappa}_1,\dots,\vec{\kappa}_n}(\chi_1,\dots,\chi_n)\equiv \frac{1}{n!} \left[\prod_{i=1}^n\hat{\varphi}^\dagger_{\vec{\kappa}_i}(\chi_i)\right]\ket{0}\bra{0}\left[\prod_{i=1}^n\hat{\varphi}_{\vec{\kappa}_i}(\chi_i)\right],
\end{equation}
which satisfy a useful orthogonality condition (note that the sum over permutations simplifies due to the symmetry in the arguments of $\hat F^{(n)}$):
\begin{align}  \label{eqn:orthogonality}
     \hat{F}^{(n)}_{\vec{\kappa}_1,\dots,\vec{\kappa}_n}(\chi_1,\dots,\chi_n) \hat{F}^{(m)}_{\vec{\kappa}'_1,\dots,\vec{\kappa}'_n}(\chi'_1,\dots,\chi'_n)&=\delta_{nm}\delta_{\vec{\kappa}_1\vec{\kappa}_1'}\cdots\delta_{\vec{\kappa}_n\vec{\kappa}_n'}\delta(\chi_1-\chi_1')\cdots\delta(\chi_n-\chi_n')\nonumber\\
     &\quad\times\hat{F}^{(n)}_{\vec{\kappa}_1,\dots,\vec{\kappa}_n}(\chi_1,\dots,\chi_n) \,.
\end{align}
The effect density $\hat{E}_\chi(\diff\chi)$ defines a POVM, and thus effect operators $\hat{E}_\chi(X)$ by
\begin{equation}\label{eqn:effectoperators}
    \hat{E}_\chi(X)\equiv \int_X\hat{E}_\chi(\diff\chi)\,,\qquad \forall X\in\mathbb{R}\,,
\end{equation}
where $X$ is any (or possibly disjoint unions of) interval(s) of $\Omega = \mathbb{R}$. An important simplification arises due to the fact that the properties of the density $\hat{E}_\chi(\diff\chi)$ are satisfied by construction also by the operator $\hat{E}_\chi(X)$. For this reason, from here on, we will focus exclusively on the density \eqref{eqn:gftpovmreal}.

First, note that the densities \eqref{eqn:gftpovmreal} are not projectors because
\begin{align}
\!\!\!\!
    \hat{E}_\chi(\diff\chi)\hat{E}_{\chi'}(\diff\chi')&=\diff\chi\diff\chi'\sum_{n=1}^\infty\sum_{\vec{\kappa}_1,\dots\vec{\kappa}_n}\int\left[\prod_{i=1}^n\diff\chi_i\right] \sum_{i,j=1}^n \frac{\delta(\chi_i-\chi)\delta(\chi_j-\chi')}{n^2}
    \hat{F}^{(n)}_{\vec{\kappa}_i}(\chi_i)\nonumber\\
    &=\diff\chi'\delta(\chi-\chi')\hat{E}_\chi(\diff\chi) \nonumber \\ & \quad
    + \diff\chi\diff\chi'\sum_{n=1}^\infty\sum_{\vec{\kappa}_1,\dots\vec{\kappa}_n} \!\! \int \! \left[\prod_{i=1}^n\diff\chi_i\right]
    \sum_{i\neq j} \frac{\delta(\chi_i-\chi)\delta(\chi_j-\chi')}{n^2}\hat{F}^{(n)}_{\vec{\kappa}_i}(\chi_i) \, , \label{eqn:notprojector}
\end{align}
using the shorthand $\hat F^{(n)}_{\vec\kappa_i}(\chi_i) = \hat{F}^{(n)}_{\vec{\kappa}_1,\dots,\vec{\kappa}_n}(\chi_1,\dots,\chi_n)$; note that the second term arises because it is possible to localize different GFT quanta at different values for $\chi$. In the QRF language, the above property characterizes $\hat{E}_\chi(\diff\chi)$ as a non-ideal QRF. Non-ideal frames are valid options for QRFs (although ideal ones, whose corresponding POVM is a projector, are simpler to study). In our case, there appears to be a trade-off between the physical interpretation of the POVM as a scalar field measurement (see equation \eqref{eqn:scalarfieldoperator} below) and the ideality of the POVM itself. We leave it to future research to determine whether a POVM can be defined that is both ideal and has a clear physical interpretation.

To gain some insights on the physical interpretation of $\hat{E}_\chi(\diff\chi)$, it is useful to study its action on $n$-particle states
\begin{equation}
    \ket{\chi_1,\vec{\kappa}_1;\dots,\chi_n,\vec{\kappa}_n}=\frac{1}{n!}\prod_{i=1}^n\hat{\varphi}_{\vec{\kappa}_i}^\dagger(\chi_i)\ket{0}.
\end{equation}
Using equation \eqref{eqn:orthogonality}, one can see that 
\begin{equation}\label{eqn:eigenstates}
    \hat{E}_\chi(\diff\chi) \ket{\chi_1,\vec{\kappa}_1;\dots,\chi_n,\vec{\kappa}_n}=\diff\chi \sum_{i=1}^n\frac{\delta(\chi_i-\chi)}{n}\ket{\chi_1,\vec{\kappa}_1;\dots,\chi_n,\vec{\kappa}_n},
\end{equation}
Since the action of the operator $\hat{E}_\chi(d\chi)$ on a generic $n$-particle state is non-zero only if at least one of the particles contained in the state has a clock quantum number equal to $\chi$, for each $n$-body sector of the GFT Fock space (except for the vacuum which it annihilates, $\hat{E}_\chi(\diff\chi)\ket{0}=0$), $\hat{E}_\chi(\diff\chi)$ localizes one of the GFT quanta at the clock value $\chi$. From equation \eqref{eqn:eigenstates}, together with the fact that the Fock vacuum combined with the $n$-particle states provide a basis for the GFT Fock space, we see that $\hat{E}_\chi(\diff\chi)\ge 0$. This, together with the $\sigma$-additivity of the measure $\diff\chi$ implies that the effect operators \eqref{eqn:effectoperators} do indeed constitute a POVM.

Notice, however, that analogously to event POVMs, the POVM is not normalized, since it does not capture the vacuum state:
\begin{equation}\label{eqn:normalization}
    \ket{0}\bra{0}+\int_\mathbb{R}\hat{E}_\chi(\diff\chi)=\one_{\mathcal{F}}\,.
\end{equation}
This reflects the fact that it is not possible to define a relational measurement if the system is in the GFT Fock vacuum. 

Crucially, in the same way as the first moment of event POVMs define coordinate observables, the first moment of the density \eqref{eqn:gftpovmreal} gives the observable (see \ref{app:sa} for a proof that this operator is essentially self-adjoint)
\begin{equation}\label{eqn:scalarfieldoperator}
    \hat{\chi}\equiv \int_\mathbb{R}\chi\hat{E}_\chi(\diff\chi)\,,
\end{equation}
to be interpreted for example as a clock observable if the field $\chi$ represents a relational clock. Indeed, using equation \eqref{eqn:eigenstates},
\begin{align}\label{eqn:scalarfieldoperatoraction}
\hat{\chi}\ket{\chi_1,\vec{\kappa}_1;\dots,\chi_n,\vec{\kappa}_n}&=\int\diff\chi  \, \chi \, \sum_i\frac{\delta(\chi_i-\chi)}{n}\ket{\chi_1,\vec{\kappa}_1;\dots,\chi_n,\vec{\kappa}_n}\nonumber\\&
=\frac{1}{n} \sum_{i=1}^n\chi_i \ket{\chi_1,\vec{\kappa}_1;\dots,\chi_n,\vec{\kappa}_n}.
\end{align}
Let us compare this equation to the action of the the second-quantized scalar field operator
\begin{equation}
    \hat{X} = \sum_{\vec{\kappa}} \int\diff\chi \,\, \chi \, \hat{\varphi}_{\vec{\kappa}}^\dagger(\chi) \, \hat{\varphi}_{\vec{\kappa}}(\chi)
\end{equation}
on $n$-particle states:
\begin{equation}
    \hat{X}\ket{\chi_1,\vec{\kappa}_1;\dots,\chi_n,\vec{\kappa}_n}=\sum_{i=1}^n\chi_i\ket{\chi_1,\vec{\kappa}_1;\dots,\chi_n,\vec{\kappa}_n}.
\end{equation}
Hence, $\hat{\chi}$ is the \textit{intensive} version of $\hat{X}$, $\hat{\chi}\sim \hat{X}/\hat{N}$. This not only provides a way of consistently defining an intensive version of the scalar field operator (which was missing in previous GFT literature), but also shows that the POVM \eqref{eqn:effectoperators} can be consistently interpreted as a frame field (which is an intensive quantity).

This statement is also corroborated by the fact that the $\hat{E}_\chi(\diff\chi)$ is \textit{covariant} with respect to the generator of relational time translations, i.e., the momentum operator
\begin{equation}
    \hat{\Pi} = -i \sum_{\vec{\kappa}} \int \diff\chi \, \hat{\varphi}_{\vec{\kappa}}^\dagger(\chi) \,\, \partial_\chi \hat{\varphi}_{\vec{\kappa}}(\chi) \, .
\end{equation}
As shown in \ref{app:cov},
\begin{equation}\label{eqn:covariancecondition}
    [\hat{E}_\chi(\diff\chi),\hat{\Pi}]=-i\partial_\chi\hat{E}_\chi(\diff\chi)\,,
\end{equation}
so defining $U(\chi)=e^{-i\hat{\Pi}\chi}$,
\begin{equation}\label{eqn:commutatormomentum}
    \hat{E}_{\chi+\chi'}(\diff\chi)=U(\chi')\hat{E}_\chi(\diff\chi)\hat{U}^\dagger(\chi')\,,
\end{equation}
which satisfies the covariance property \eqref{eqn:covariance}.

Importantly, this procedure can be straightforwardly generalized for a full \textit{frame} made of $m>1$ minimally-coupled massless scalar fields. In this case, \eqref{eqn:gftpovmreal} becomes
\begin{equation}
    \hat{E}_{\boldsymbol{\chi}}(\diff\boldsymbol{\chi}) = \diff^m\chi \sum_{n=1}^\infty \sum_{\vec{\kappa}_1,\dots\vec{\kappa}_n} \!\! \int \!\! \left[\prod_{i=1}^n \diff^m\chi_i \diff^\ell\phi_i \right] \! \frac{1}{n} \sum_{i=1}^n\delta^{(m)}(\boldsymbol{\chi}_i-\boldsymbol{\chi}) \hat{F}^{(n)}_{\vec{\kappa}_1,\dots,\vec{\kappa}_n}(\boldsymbol{\chi}_1,\vec{\phi}_1,\dots,\boldsymbol{\chi}_n,\vec{\phi}_n),
\end{equation}
where $\boldsymbol{\chi}=\{\chi^1,\dots,\chi^m\}$ and $\vec{\phi}=\{\phi_1,\dots,\phi_\ell\}$ collectively denote the frame and other scalar fields, respectively. It is easy to see that the effect operators constructed out of this density constitute a (non-normalized) \textit{frame} POVM, as they are additive and positive, and further are covariant with respect to the momenta 
\begin{equation}
    \hat{\Pi}^a\equiv -i\sum_{\vec{\kappa}}\int\diff^m\chi\diff^\ell\phi\,\hat{\varphi}_{\vec{\kappa}}^\dagger(\boldsymbol{\chi},\vec{\phi})\partial_{\chi^a}\hat{\varphi}_{\vec{\kappa}}(\boldsymbol{\chi},\vec{\phi})\,,
\end{equation}
while the first moments 
\begin{equation}
    \hat{\chi}^a\equiv \int_{\mathbb{R}^m}\chi^a\hat{E}_{\boldsymbol{\chi}}(\diff\boldsymbol{\chi})
\end{equation}
provide a consistent definition of intensive frame field components. (These other scalar fields could for example play the role of relational rods, although in that case the GFT action will depend on the rod fields differently than on the clock field, see \cite{Jercher:2023kfr, Jercher:2023nxa} for one such construction.) Thus, for the sake of simplicity, from here on we will focus on a single frame component, that we will often refer to as a clock.

\subsection{Relational observables in GFT}
\label{sec:relationalobservablesgft}

Constructing relational observables in GFT is both conceptually important, and also critical for practical reasons: due to the complete absence of continuum spacetime notions, any correspondence with continuum physics within GFT can only be established in a relational framework \cite{Oriti:2024}; this is why this approach has thus far been the primary strategy employed to derive cosmological physics from GFT.

The absence of continuum spacetime structures is also why relational observables cannot be directly constructed using the single-integral and canonical procedures outlined in \eqref{eqn:singleintegral} and \eqref{eqn:canonical}. Instead, relational observables in GFT have so far been developed following two main approaches.

A first approach consists in sharply localizing second-quantized operators in the frame domain \cite{Oriti:2016qtz, Gielen:2016dss, Gielen:2017eco, Gielen:2018xph, Gielen:2018fqv, Gerhardt:2018byq}. For instance, for the number operator introduced in equation \eqref{eqn:numberoperator}, and in the case of a single clock, this procedure would amount to
\begin{equation}\label{eqn:relobservablesold}
        \hat{N}=\sum_{\vec{\kappa}}\int\diff\chi ~ \varphi^\dagger_{\vec{\kappa}}(\chi)\hat{\varphi}_{\vec{\kappa}}(\chi)\quad\longrightarrow\quad \hat{N}(\chi)\equiv\sum_{\vec{\kappa}}\varphi^\dagger_{\vec{\kappa}}(\chi)\hat{\varphi}_{\vec{\kappa}}(\chi)\,.
\end{equation}
Although apparently straightforward, as noted in \cite{Marchetti:2020qsq}, this prescription can lead to several issues, particularly when using a perspective-neutral description of the GFT Fock space characterized by the commutation relations \eqref{eqn:commutationrelationsk} in which case the resulting operator is distributional and must be regularized by integrating (or smearing) over a small interval of $\chi$. This technical problem is avoided by postulating a perspective-reduced Fock space \cite{Wilson-Ewing:2018mrp, Gielen:2019nyq, Gielen:2019kae, Gielen:2020fgi, Gielen:2023szb}, which, however, relies on fixing a clock before quantization, and therefore ignores all quantum fluctuations in the clock field as well as its conjugate momentum; these points are further discussed in Sec.~\ref{sec:deparametrized}.

A second approach that has previously been used involves constructing \textit{effective} relational observables as averages over appropriate states \cite{Marchetti:2020umh,Marchetti:2020qsq,Gielen:2021vdd,Marchetti:2021gcv,Jercher:2021bie,Jercher:2023kfr,Jercher:2023nxa}. An example of these states is provided by coherent states, which are localized (peaked) around certain arbitrary configurations of the frame \cite{Marchetti:2020qsq}. Although this addresses some of the issues associated with the first approach, the dependence on specific (and typically collective and classical) states complicates the study of quantum correlations of these observables at different frame values, and is only applicable to certain classes of states.

In this work, instead, we propose to use an operative definition of relational observables based on \eqref{eqn:operativerelational}: for any operator $\hat{O}$ acting on the GFT Fock space $\mathcal{F}$, its corresponding relational observable (density) is defined as the Jordan product
\begin{equation}\label{eqn:relobsgft}
    \hat{O}_\chi(\diff\chi)\equiv \left\{\hat{O},\hat{E}_\chi(\diff\chi)\right\},
\end{equation}
where $\{ a , b\}=(ab+ba)/2$ represents the anticommutator, introduced in order to avoid ordering ambiguities; note that this map is unital. We note that other choices are also possible; for example, alternate definitions could be $\hat{O}'_\chi(\diff\chi) \equiv \sqrt{\hat{E}_\chi(\diff\chi)} \, \hat{O} \, \sqrt{\hat{E}_\chi(\diff\chi)}$, or else (viewing the effect densities as nearly being projectors) $\hat{O}''_\chi(\diff\chi) \equiv \hat{E}_\chi(\diff\chi) \, \hat{O} \, \hat{E}_\chi(\diff\chi)$, although these lead to more complicated calculations and to a less straightforward connection with the definition of relational observables used in the QRF context \eqref{eqn:operativerelational}. For both of these reasons, we will use the definition \eqref{eqn:relobsgft} for the remainder of the paper.

Although this definition is motivated in part by trying to adapt \eqref{eqn:operativerelational} to GFT, it is not possible to directly apply that equation to GFT due to important differences in the Hilbert spaces and the role of gauge symmetries in the theories. For this reason, the definition \eqref{eqn:relobsgft} technically differs from \eqref{eqn:operativerelational} in two important aspects.

First, $\hat{O}^{(S)}_R$ in equation \eqref{eqn:operativerelational} is obtained by tensoring the partial observables $\hat{O}^{(S)}$ and $\ket{\phi(g)}_R\bra{\phi(g)}_R$. This specific structure follows from the assumption that the kinematical Hilbert space of the theory also has a tensor structure, $\mathcal{H}_{\text{kin}}=\mathcal{H}_R\otimes \mathcal{H}_S$, its two factors corresponding to the reference frame space and the system space, respectively. The GFT Fock space does not satisfy this assumption, as both the frame (e.g., a clock scalar field) and the system (e.g., geometry) are expected to emerge from the dynamics of the same GFT quanta. Thus, both $\hat{O}$ and $\hat{E}_\chi(\diff\chi)$ act on the entirety of the GFT Fock space $\mathcal{F}$.

Second, equation \eqref{eqn:relobsgft} does not include a group averaging. The group averaging was introduced in equation \eqref{eqn:operativerelational} to ensure that the resulting relational observable commutes with the (quantum) constraint generating the gauge symmetry group $G$ (i.e., that it is $G$-gauge invariant). In gravitational theories, this symmetry group typically consists of (small) diffeomorphisms, and relational observables need to be invariant under the action of generators of these transformations. In GFT, the situation is markedly different. Since GFT abandons traditional spacetime structures, there is no reason to expect constraints associated with emergent diffeomorphism invariance. Nonetheless, this symmetry is anticipated to affect the structure of the GFT, particularly its dynamics. This is suggested by the results of \cite{Marchetti:2024tjq}, where a GFT quantization of a background-independent model without a Hamiltonian constraint (known as the Husain-Kucha\v{r} model \cite{Husain:1990}) resulted in a non-interacting GFT. Therefore, we expect that as long as equation \eqref{eqn:relobsgft} is evaluated on shell, the resulting observable will be gauge-invariant in the continuum limit. Note that by evaluating equation \eqref{eqn:relobsgft} on shell, one may also expect, based on results established in the context of QRFs (see, e.g., \cite{Hoehn:2019fsy, delaHamette:2021oex}), that the observable defined is equivalent to a quantum gauge-reduced one.

It is useful to write equation \eqref{eqn:relobsgft} more explicitly, specializing to the case of a one-body operator (partial observable) of the form
\begin{equation}
    \hat{O}=\sum_{\vec{\kappa}_1,\vec{\kappa}_2}\int\diff\chi_1\diff\chi_2 ~ O_{\vec{\kappa}_1,\vec{\kappa}_2}(\chi_1,\chi_2) \, \hat{\varphi}^\dagger_{\vec{\kappa}_1}(\chi_1) \, \hat{\varphi}_{\vec{\kappa}_2}(\chi_2)\,.
\end{equation}
Moreover, for the sake of simplicity, we will assume its kernel to be symmetric with respect to its arguments: $O_{\vec{\kappa}_1,\vec{\kappa}_2}(\chi_1,\chi_2)=O_{\vec{\kappa}_2,\vec{\kappa}_1}(\chi_1,\chi_2)=O_{\vec{\kappa}_1,\vec{\kappa}_2}(\chi_2,\chi_1)$. We make this assumption to simplify calculations, and also because several important operators, including the number and volume operators, satisfy this symmetry; however, in general this symmetry will not be satisfied for all GFT operators. Under this assumption, $[\hat{O},\hat{E}_\chi(\diff\chi)]=0$, so $\hat{O}_\chi(\diff\chi)=\hat{O}\hat{E}_\chi(\diff\chi)$, and, by using equation \eqref{eqn:commutatorfa}, $\hat{O}_\chi$ takes the form
\begin{align}\label{eqn:relobsdifferentform}
    \hat{O}_\chi(\diff\chi) &= \diff\chi \sum_{n=1}^\infty \sum_{\vec{\kappa}_1,\dots\vec{\kappa}_n} \int \left[\prod_{i=1}^n\diff\chi_i\right] \frac{1}{n^2} \sum_{j=1}^n\delta(\chi_j-\chi) \nonumber\\
    &\qquad\times \sum_{\ell=1}^n\sum_{\vec{\kappa}'} \int\diff\chi' \, O_{\vec{\kappa}',\vec{\kappa}_\ell}(\chi',\chi_\ell) \, \hat{\varphi}^\dagger_{\vec{\kappa}'}(\chi') \, \hat{F}^{(n-1)}_{[j\neq \ell]} \, \hat{\varphi}_{\vec{\kappa}_\ell}(\chi_\ell) \,,
\end{align} 
where $\hat{F}^{(n-1)}_{[j\neq \ell]}$ is defined below \eqref{eqn:commutatorfa}. Note that in this case, the map $\chi: \hat{O}\mapsto \hat{O}_\chi$ is positive, but it is not completely positive. However, it is important to stress that in this context, complete positivity is less critical than in standard quantum mechanical systems: adding an additional (sub)system does not enlarge the GFT Fock space in the usual tensor-product sense, but rather extends the domain over which the GFT field is defined. If $\hat O$ is not symmetric, it is necessary to average this with a second contribution obtained by exchanging the $\vec \kappa', \chi'$ and $\vec \kappa_\ell, \chi_\ell$ labels on $O_{\vec{\kappa}',\vec{\kappa}_\ell}(\chi',\chi_\ell)$ as well as between the operators $\hat{\varphi}^\dagger_{\vec{\kappa}'}(\chi')$ and $\hat{\varphi}_{\vec{\kappa}_\ell}(\chi_\ell)$.

\section{Comparison with previous approaches}
\label{sec:comparison}

On a generic one-particle state $\ket{\vec{\kappa}_1,\chi_1}$, it follows from \eqref{eqn:relobsgft}, or equivalently \eqref{eqn:relobsdifferentform} since all higher-body operators vanish on a one-particle state, that
\begin{equation}
    \hat{O}_\chi(\diff\chi)\ket{\vec{\kappa}_1,\chi_1}=\diff\chi\, \hat{O}(\chi) \delta(\chi - \chi_1) \ket{\vec{\kappa}_1,\chi_1}.
\end{equation}
Thus, within the one-particle sector of the GFT Fock space, the relational observables \eqref{eqn:relobsgft} exactly match the sharply localized observables defined in \eqref{eqn:relobservablesold}. In general, however, $\hat{O}_\chi(\diff\chi)$ differs substantially from the definition of sharply localized observables. This is due to the fact that in general, each GFT quantum carries its own \qmarks{clock value} \cite{Marchetti:2020qsq, Kotecha:2018gof}: different GFT quanta need not be synchronized. In other words, the difference between \eqref{eqn:relobsgft} and \eqref{eqn:relobservablesold} is due to the intrinsically multi-fingered nature of time of the (clock-neutral) GFT Fock space. We will explore this issue further in Sec.~\ref{sec:deparametrized}.

Even though these two definitions of relational observables differ at the microscopic, quantum level, they may still agree at the macroscopic level, i.e., when averaged over certain appropriate semi-classical states.

\subsection{Expectation values on coherent states}
\label{sec:coherentstates}

In particular, it is important to check the robustness of previous results obtained in the context of cosmology using the sharply localized observables to ensure they continue to hold (up to possible small corrections) if POVM-based relational observables are used instead. To this purpose, since previous works focused on one-body operators $O$ that are symmetric in the $SU(2)$ and matter field labels, we will now compute the expectation value of \eqref{eqn:relobsdifferentform} on coherent states.

Coherent states $\ket{\sigma}$ are eigenstates of the operator $\hat{\varphi}_{\vec{\kappa}(\chi)}$ and are defined in \eqref{eqn:coherentstates}. Using the indistinguishability of bosonic particles, together with equation \eqref{eqn:relobsdifferentform}, the expectation value of $\hat{O}_\chi(\diff\chi)$ on $\ket{\sigma}$ is
\begin{align}\label{eqn:expvaluecoherent1}
    \langle\hat{O}_\chi(\diff\chi)\rangle_\sigma&=\vert\mathcal{N}_\sigma\vert^2\diff\chi\sum_{n=1}^\infty\frac{1}{n!}\sum_{\vec{\kappa}_1,\dots\vec{\kappa}_n}\int\left[\prod_{i=1}^n\diff\chi_i\right] \left(\sum_{i=1}^n\delta(\chi_i-\chi) \right) \nonumber\\
    &\quad\times \sum_{\vec{\kappa}'}\int\diff\chi'\,O_{\vec{\kappa}',\vec{\kappa}_1}(\chi',\chi_1)\bar{\sigma}_{\vec{\kappa}'}(\chi')\sigma_{\vec{\kappa}_1}(\chi_1)\left[\prod_{i=2}^n\vert\sigma_{\vec{\kappa}_i}(\chi_i)\vert^2\right].
\end{align}
It is convenient to separate the term where the delta function acts on the arguments of the kernel $O_{\vec{\kappa}_1,\vec{\kappa}'}(\chi_1,\chi')$. By introducing
\begin{subequations}\label{eqn:notation1}
\begin{align}
    N_\sigma&\equiv \sum_{\vec{\kappa}}\int\diff\chi\vert\sigma_{\vec{\kappa}}(\chi)\vert^2\,,\\
    N_\sigma(\chi)&\equiv \sum_{\vec{\kappa}}\vert\sigma_{\vec{\kappa}}(\chi)\vert^2\,,\\
    O_\sigma(\chi)&\equiv \sum_{\vec{\kappa},\vec{\kappa}'}\int\diff\chi'\,O_{\vec{\kappa}',\vec{\kappa}}(\chi',\chi)\bar{\sigma}_{\vec{\kappa}'}(\chi')\sigma_{\vec{\kappa}}(\chi)\,,
\end{align}
\end{subequations}
the expectation value $\langle\hat{O}_\chi(\diff\chi)\rangle_\sigma$ can be rewritten as (see \ref{app:coherent} for a derivation)
\begin{equation}\label{eqn:expvaluegenericcoherent2}
    \langle\hat{O}_\chi(\diff\chi)\rangle_\sigma=\diff\chi\left\{\frac{1-\vert \mathcal{N}_\sigma\vert^2}{N_\sigma}\left[O_\sigma(\chi)-\langle\hat{O}\rangle_\sigma\frac{N_\sigma(\chi)}{N_\sigma}\right]+\langle\hat{O}\rangle_\sigma\frac{N_\sigma(\chi)}{N_\sigma}\right\},
\end{equation}
where $\langle O \rangle_\sigma = \langle \sigma | O | \sigma \rangle$. Note that for highly quantum coherent states, i.e., states with small number of quanta $N_\sigma \ll 1$, one has $1-\vert\mathcal{N}_\sigma\vert^2\simeq N_\sigma$, so
\begin{equation}\label{eqn:lowdensity}  \langle\hat{O}_\chi(\diff\chi)\rangle_\sigma\underset{N_\sigma\ll 1}{\approx}\diff\chi\,O_\sigma(\chi)\,.
\end{equation}
In this limit, the expectation value of the relational observable (density) $\hat{O}_\chi(\diff\chi)$ on the coherent states \eqref{eqn:coherentstates} reduces to the expectation value of the sharply localized observable \eqref{eqn:relobservablesold}. Intuitively, this is because these states rarely contain more than one particle, and thus they are effectively synchronized. On the other hand, when $N_\sigma\gg 1$, the collective nature of the state is more manifest and in this case these two expectation values will not match. Indeed, in the opposite limit of large number of quanta $N_\sigma\gg 1$, then $\mathcal{N}_\sigma \to 0$ and
\begin{equation}\label{eqn:highdensity}  \langle\hat{O}_\chi(\diff\chi)\rangle_\sigma\underset{N_\sigma\gg 1}{\approx}\frac{\diff\chi}{N_\sigma}\left[O_\sigma(\chi)+\langle\hat{O}\rangle_\sigma N_\sigma(\chi)\right],
\end{equation}
where the collective properties of the state affect the behavior of the relational observable average by introducing a term proportional to $N_\sigma(\chi)$. 

We can now apply \eqref{eqn:expvaluegenericcoherent2} to specific examples of observables of physical interest:
\begin{description}
    \item[Number operator.]The case of the number operator is particularly simple, and yet extremely important for physical applications, since it has been shown that the number operator controls the continuum and classical limit in cosmological applications of GFT \cite{Marchetti:2020qsq,Gielen:2019kae}. For the number operator,  $O^{(N)}_{\vec{\kappa}_1,\vec{\kappa}_2}(\chi_1,\chi_2)=\delta(\chi_1-\chi_2)\delta_{\vec{\kappa}_2,\vec{\kappa}_2}$. As a consequence $O^{(N)}_\sigma(\chi)=N_\sigma(\chi)$ and $\langle\hat{O}^{(N)}\rangle_\sigma=N_\sigma$. The two terms in square brackets in \eqref{eqn:expvaluegenericcoherent2} therefore cancel, and one is left with
\begin{equation}
    \langle \hat{N}_\chi(\diff\chi)\rangle_\sigma=\diff\chi\,N_\sigma(\chi)\,,
\end{equation}
which exactly matches the expectation value of the corresponding sharply localized observable \eqref{eqn:relobservablesold}.
\item[Scalar field operator.] Let us now consider the case of the second-quantized scalar field operator, $O^{(X)}_{\vec{\kappa}_1,\vec{\kappa}_2}(\chi_1,\chi_2)=\chi_1\delta(\chi_1-\chi_2)\delta_{\vec{\kappa}_2,\vec{\kappa}_2}$, so that $O^{(X)}_\sigma(\chi)\equiv X_\sigma(\chi)=\chi N_\sigma(\chi)$. If $\vert\sigma_{\vec{\kappa}}(\chi)\vert^2$ is symmetric in $\chi$, then $\langle\hat{X}\rangle_\sigma=0$, so that
\begin{equation}
    \langle\hat{X}_\chi(\diff\chi)\rangle_\sigma=\diff\chi\frac{1-\vert \mathcal{N_\sigma}\vert^2}{N_\sigma}\chi N_\sigma(\chi)=\diff\chi\frac{1-\vert \mathcal{N_\sigma}\vert^2}{N_\sigma}X_\sigma(\chi)\,.
\end{equation}
One can attempt to extract intensive information out of the above expectation value by dividing it by $N_\sigma(\chi)$. The result is proportional to $\chi\diff\chi$, as expected. Notice that even when $\vert\sigma\vert^2$ is not symmetric in $\chi$, since $\partial_\chi \langle \hat X \rangle_\sigma = 0$ it follows that
\begin{equation}
    \frac{\diff}{\diff\chi }\frac{\langle\hat{X}_\chi(\diff\chi)\rangle_\sigma}{N_\sigma(\chi)}=\frac{1-\vert \mathcal{N}\vert^2}{N_\sigma}=\text{const.}\,,
\end{equation}
as one would expect for monotonic clock.
\item[Volume operator.]
Finally, we consider the volume operator $\hat{V}$, defined by $O^{(V)}_{\vec{\kappa}_1,\vec{\kappa}_2}(\chi_1,\chi_2)=V_{\vec{\kappa}_1}\delta(\chi_1-\chi_2)\delta_{\vec{\kappa}_2,\vec{\kappa}_2}$. In general, the expression for its expectation value is quite involved. However, in the case of GFT condensate cosmology, condensates satisfying the conditions $\sigma_{\vec{\kappa}}=\sigma_{j_o}$ (where the condensate is dominated by quanta of geometry whose spins are all $j=j_o$, and are eigenstates of maximal volume given these spins) are of particular interest \cite{Oriti:2016qtz, Gielen:2016uft}, and in this case $O^{(V)}_{\sigma_{j_o}}(\chi)\equiv V_{\sigma_{j_o}}(\chi) = V_{j_o} N_{\sigma_{j_o}}(\chi)$. This implies that, since this is a rescaling of the number operator by $V_{j_o}$, the square brackets in \eqref{eqn:expvaluegenericcoherent2} again cancel and
\begin{equation}
    \langle\hat{V}_\chi\rangle_{\sigma_{j_o}}=\diff\chi V_{j_o}N_{\sigma_{j_o}}(\chi)=\diff\chi V_{\sigma_{j_o}}(\chi)\,,
\end{equation}
which exactly matches the expectation value of the sharply localized volume operator \eqref{eqn:relobservablesold} \cite{Oriti:2016qtz}. Crucially, this confirms that the results obtained in the cosmological context by using \eqref{eqn:relobservablesold} to construct a sharply localized volume observable apply to the relational observable $\hat{V}_\chi(\diff\chi)$ as well (although only for single-spin condensate states $\sigma_{j_o}$; in contrast, there could be non-negligible differences for other types of condensate states).
\end{description}

\subsection{Sharp localization and measurement resolution}
\label{sec:effectiverelobs}

Sharply localizing observables in GFT with respect to a physical clock generally produces significant quantum fluctuations in the clock momentum observable \cite{Marchetti:2020umh, Marchetti:2020qsq}. Given that the clock effect density \eqref{eqn:gftpovmreal} involves a sharp localization within the clock domain, one can expect similar behavior in this case. To see this, consider the relational momentum operator
\begin{equation}
    \hat{\Pi}_\chi=\left\{\hat{E}_\chi(\diff\chi) ,\hat{\Pi}\right\},
\end{equation}
that satisfies the following commutation relation with $\hat{E}_\chi(\diff\chi)$
\begin{equation}\label{eqn:commutatorirelationalpi}
    [\hat{E}_\chi(\diff\chi),\hat{\Pi}_{\chi'}]= - \frac{i}{2} \left( \left( \partial_\chi\hat{E}_\chi(\diff\chi) \right) \cdot \hat{E}_{\chi'}(\diff\chi')
    + \hat{E}_{\chi'}(\diff\chi') \cdot \partial_\chi\hat{E}_\chi(\diff\chi)
    \right) ,
\end{equation}
since $\hat{E}_\chi(\diff\chi)$ is $\Pi$-covariant. Note that the right-hand side of this equation includes terms of the form $\partial_\chi [\delta (\chi - \chi_i)] \cdot \delta (\chi' - \chi_j)$ where $\chi_i, \chi_j$ are the fields summed over. In the case $\chi=\chi'$, one such term will have the `diagonal' form $\partial_\chi [\delta (\chi - \chi_i)] \cdot \delta (\chi - \chi_i)$, which is proportional to $\delta(0)$, and divergent (there is no such divergence for $\chi \neq \chi'$). This directly affects the behavior of quantum fluctuations on the relational momentum observable $\hat{\Pi}_\chi$. Since the uncertainty relation for any two observables $\hat{A}$ and $\hat{B}$ is \cite{Puri2001}
\begin{equation}\label{eqn:uncertainty}
    \Delta\hat{A}^2\Delta\hat{B}^2\ge\frac{1}{4}\left[\langle\hat{F}\rangle^2+\langle\hat{C}\rangle^2\right],
\end{equation}
where $\Delta \hat{O}\equiv \langle(\hat{O}-\langle \hat{O}\rangle)^2\rangle$ is the variance of $\hat{O}$, and where
\begin{equation}
    \hat{F}=\hat{A}\hat{B}+\hat{B}\hat{A}-2\langle\hat{A}\rangle\langle\hat{B}\rangle\,,\qquad i\hat{C}=[\hat{A},\hat{B}]\,,
\end{equation}
it is clear that there is a divergent term contributing to $\Delta \hat E_\chi(\diff\chi)^2 \, \Delta \Pi_\chi^2$.

The divergent term is not strictly positive, and vanishes for any state that has no support at the specific value of $\chi$ being considered. For example, applying \eqref{eqn:eigenstates} to a generic $n$-particle state,
\begin{align}
    \partial_\chi\hat{E}_\chi(\diff\chi)\ket{\chi_1,\vec{\kappa}_1;\dots,\chi_n,\vec{\kappa}_n}&=\diff\chi \sum_{i=1}^n\frac{\partial_\chi\delta(\chi_i-\chi)}{n}\ket{\chi_1,\vec{\kappa}_1;\dots,\chi_n,\vec{\kappa}_n}
    ,
\end{align}
the right-hand side vanishes for any state where $\chi_i \neq \chi$, and the divergent term does not arise. Similarly for coherent states,
\begin{equation}
    \ket{\sigma}=\mathcal{N}_\sigma\sum_{n=0}^\infty\sum_{\vec{\kappa}_1,\dots,\vec{\kappa}_n}\int\left[\prod_{i=1}^n\diff\chi_i\right]\frac{\sigma_{\vec{\kappa}_1}(\chi_1)\cdots \sigma_{\vec{\kappa}_n}(\chi_n)}{n!}\ket{\chi_1,\vec{\kappa}_1;\dots,\chi_n,\vec{\kappa}_n},
\end{equation}
so $\partial_\chi\hat{E}_\chi(\diff\chi)\ket{\sigma}$ is non-vanishing only if $\sum_{\vec{\kappa}}\sigma_{\vec{\kappa}}(\chi')\hat{\varphi}^\dagger_\kappa(\chi')$ has support on $\chi$. Since condensate states are commonly used to extract continuum physics from GFT, a divergent clock momentum variance on these states may well have important implications for the resulting emergent continuum physics.

Note that in principle, a similarly divergent term is also present in the quantity $\hat{F}$, as $\hat{F}$ in this case contains a term of the form $\hat{\Pi}_\chi\hat{E}_\chi=\hat{\Pi}\hat{E}_\chi^2$ that is analogous to that on the right-hand side of equation \eqref{eqn:commutatorirelationalpi}. However, it is easy to see that since $n$-particle states are eigenvalues of $\hat{E}_\chi(\diff\chi)$, the expectation value of $\hat{F}$ on those states vanishes. The same argument applies to operators $\hat{O}_\chi$ that commute with $\hat{E}_\chi(\diff\chi)$, meaning that for such operators it is always possible to find states with arbitrarily small variance. Examples of these operators are the volume operator, the number operator, and the scalar field operator discussed in Sec.~\ref{sec:coherentstates}. 

As mentioned above, large quantum fluctuations in the clock momentum are inherently due to the sharp localization provided by the effect density \eqref{eqn:gftpovmreal}. Such perfect time localization can clearly only be an idealization of a physical measurement with uncertainty $\epsilon$ (in the limit in which $\epsilon\to 0$). One possible way to avoid this divergence would be to integrate $\hat E_\chi(\diff \chi)$ over a small range of $\chi$ to obtain a suitably smeared operator; alternatively, to describe a more realistic measurement process characterized by a variance $\epsilon$, another possibility is to replace the $\delta$-function in \eqref{eqn:gftpovmreal} with \cite{Distler:2012eh}
\begin{equation}\label{eqn:gaussian}
    \eta_{\epsilon}(x)=\frac{1}{\sqrt{2\pi\epsilon}}\exp[-x^2/(2\epsilon)]\,,
\end{equation}
and define a realistic effect density as
\begin{equation}
    \hat{\mathcal{E}}_{\chi;\epsilon}(\diff\chi)\equiv\diff\chi\sum_{n=1}^\infty\sum_{\vec{\kappa}_1,\dots\vec{\kappa}_n}\int\left[\prod_{i=1}^n\diff\chi_i\right]\frac{\sum_{i=1}^n\eta_\epsilon(\chi_i-\chi)}{n}\hat{F}^{(n)}_{\vec{\kappa}_1,\dots,\vec{\kappa}_n}(\chi_1,\dots,\chi_n) \,.
\end{equation}
Note that $\hat{\mathcal{E}}_{\chi;\epsilon}(\diff\chi)$ also exactly satisfies the counterparts of \eqref{eqn:normalization} and \eqref{eqn:covariancecondition}
\begin{equation}
    \ket{0}\bra{0}+\int_\mathbb{R}\hat{\mathcal{E}}_{\chi;\epsilon}(\diff\chi)=\one\,, \qquad \qquad
    [\hat{\mathcal{E}}_{\chi;\epsilon}(\diff\chi),\hat{\Pi}]=-i\partial_\chi\hat{\mathcal{E}}_{\chi;\epsilon}(\diff\chi)\,,
\end{equation}
and it also satisfies 
\begin{equation}\label{eqn:regularizedcommutator}
    [\hat{\mathcal{E}}_{\chi;\epsilon}(\diff\chi),\hat{\Pi}_{\chi'}]= - \frac{i}{2} \left( \left( \partial_\chi\hat{\mathcal{E}}_{\chi;\epsilon}(\diff\chi) \right) \cdot \hat{\mathcal{E}}_{\chi';\epsilon}(\diff\chi')
    + \hat{\mathcal{E}}_{\chi';\epsilon}(\diff\chi') \cdot \partial_\chi\hat{\mathcal{E}}_{\chi;\epsilon}(\diff\chi)
    \right) .
\end{equation}
Similarly to $\hat E_\chi(\diff \chi)$, the effect density $\hat{\mathcal{E}}_{\chi;\epsilon}(\diff\chi)$ is not a projector since, to lowest order in $\epsilon$,
\begin{align}
    \hat{\mathcal{E}}_{\chi;\epsilon}(\diff\chi)\hat{\mathcal{E}}_{\chi';\epsilon}(\diff\chi')&\simeq \diff\chi'\eta_\epsilon(\chi-\chi')\hat{\mathcal{E}}_{\chi;\epsilon}(\diff\chi)+ \diff\chi\diff\chi'\sum_{n=1}^\infty\sum_{\vec{\kappa}_1,\dots\vec{\kappa}_n}\int\left[\prod_{i=1}^n\diff\chi_i\right]\nonumber\\
    &\quad\times\frac{\sum_{i\neq j}\eta_\epsilon(\chi_i-\chi)\eta_\epsilon(\chi_j-\chi')}{n^2}\hat{F}^{(n)}_{\vec{\kappa}_1,\dots,\vec{\kappa}_n}(\chi_1,\dots,\chi_n)\,.
\end{align}
This implies, in particular, that the previously divergent term appearing at the right-hand-side of \eqref{eqn:regularizedcommutator} is now regularized to $-i\diff\chi(2\pi\epsilon)^{-1/2}\partial_\chi \hat{\mathcal{E}}_{\chi;\epsilon}(\diff\chi)$. As expected, the smaller the intrinsic uncertainty $\epsilon$ associated with the clock measurement is, the larger the quantum fluctuations on the clock momentum must be.

Note that if, instead of $\hat{E}_{\chi}(\diff\chi)$, it is the effect density $\hat{\mathcal{E}}_{\chi;\epsilon}(\diff\chi)$ that is used to define the scalar field operator $\hat{\chi}_\epsilon$ and the relational observables $\hat{O}_{\chi;\epsilon}$ in \eqref{eqn:scalarfieldoperator} and \eqref{eqn:relobsgft} respectively, then the results \eqref{eqn:scalarfieldoperatoraction} and \eqref{eqn:expvaluegenericcoherent2} continue to hold to lowest order in $\epsilon$. In particular, when evaluating this new definition of relational observables to coherent states, up to corrections of order $\epsilon$,
\begin{equation}
   \langle \hat{N}_{\chi;\epsilon}(\diff\chi)\rangle_\sigma\simeq \diff\chi\,N_\sigma(\chi)\,,\qquad
   \langle\hat{V}_{\chi;\epsilon}\rangle_{\sigma_{j_o}}\simeq \diff\chi \, V_{j_o}N_\sigma(\chi)=\diff\chi \, V_{\sigma_{j_o}}(\chi)\,,
\end{equation}
showing that (to leading order) the density $\hat{\mathcal{E}}_{\chi;\epsilon}(\diff\chi)$ produces the same number and volume effective relational observables defined for coherent peaked states \cite{Marchetti:2020qsq}. Note, however, that this family of coherent peaked states allow small quantum fluctuations (in appropriate regimes) on both the clock observable and its momentum by including a non-trivial phase in the peaking function \eqref{eqn:gaussian} \cite{Marchetti:2020qsq}. An important difference is that the perspective adopted here is state-independent, as the quantum measurement of the clock is captured through the density $\hat{\mathcal{E}}_{\chi;\epsilon}(\diff\chi)$ only, while (when appropriate) semi-classicality requirements should be imposed on states.

\subsection{Perspective-reduced approach}
\label{sec:deparametrized}

As already discussed, the relational observables constructed from \eqref{eqn:relobsgft} differ from the sharply localized observables in \eqref{eqn:relobservablesold} used in previous approaches, essentially due to the intrinsically multi-fingered nature of time in the perspective neutral GFT Fock space. In this section, we confirm this intuition by showing that ultra-localized versions of such observables can be constructed via the operative POVM prescription \eqref{eqn:gftpovmreal}. In doing so, we also show how a perspective-reduced Fock space can be constructed by projection onto a particular subspace of $\mathcal{F}$.

To do this, we introduce
\begin{equation}
    \hat{P}_\chi=\ket{0}\bra{0}+\sum_n\sum_{\kappa_1,\dots,\kappa_n}\delta(0)^{-n}\hat{F}^{(n)}_{\vec{\kappa}_1,\dots,\vec{\kappa}_n}(\chi)\,,
\end{equation}
where $\hat{F}_{\kappa_1,\dots,\kappa_n}(\chi)\equiv \hat{F}_{\kappa_1,\dots,\kappa_n}(\chi,\dots,\chi)$. As seen in Sec.~\ref{sec:effectiverelobs}, the product of two localizing functions at the same clock value is in general divergent and needs to be regularized, for example by introducing a width $\epsilon$ to the measurement of $\chi$. Here, for the sake of simplicity, we simply include appropriate factors of $\delta(0)$ to guarantee the regularity of $\hat{P}_\chi$, but this should be understood as a shorthand for an appropriate regularization of the operator.

It is easy to see from \eqref{eqn:orthogonality} that $\hat{P}_\chi$ is a projector,
\begin{equation}
\hat P_\chi^2 = \hat P_\chi \,.
\end{equation}
The projected space is a \qmarks{synchronized Fock space}, as a generic $n$-body state $\ket{\vec{\kappa}_1,\chi_1,\dots,\vec{\kappa}_n,\chi_n}$ is projected onto
\begin{equation}
    \hat{P}_\chi\ket{\vec{\kappa}_1,\chi_1,\dots,\vec{\kappa}_n,\chi_n}=\frac{\prod_{i=1}^n\delta(\chi_i-\chi)}{\delta(0)^n}\ket{\vec{\kappa}_1,\chi,\dots,\vec{\kappa}_n,\chi}.
\end{equation}
Note, in particular, that all $n$-particle states with $\chi_i\neq \chi$ for any $i=1,\dots,n$, belong to the kernel of this projector. The projection $\hat{P}_\chi\mathcal{F}_{\text{GFT}}=\mathcal{F}_\chi$ thus only contains synchronized $n$-body states such as $\ket{\vec{\kappa}_1,\chi,\dots,\vec{\kappa}_n,\chi}$, and can be constructed out of the field operators
\begin{equation}\label{eqn:projectedfieldops}
    \hat{\tilde{\varphi}}_{\vec{\kappa}}(\chi)\equiv\frac{ \hat{P}_\chi\hat{\varphi}_{\vec{\kappa}}(\chi)\hat{P}_\chi}{\sqrt{\delta(0)}}\,,\qquad \hat{\tilde{\varphi}}^\dagger_{\vec{\kappa}}(\chi)\equiv \frac{\hat{P}_\chi\hat{\varphi}^\dagger_{\vec{\kappa}}(\chi)\hat{P}_\chi}{\sqrt{\delta(0)}}\,,
\end{equation}
so that
\begin{equation}\label{eqn:sametimecommutationrelations}
    [\hat{\tilde{\varphi}}_{\vec{\kappa}}(\chi),\hat{\tilde{\varphi}}^\dagger_{\vec{\kappa}'}(\chi)]= \delta_{\vec{\kappa}\vec{\kappa}'}P_\chi\,.
\end{equation}
These are the same-time commutation relations assumed in the perspective-reduced framework of \cite{Wilson-Ewing:2018mrp, Gielen:2019nyq, Gielen:2019kae, Gielen:2020fgi, Gielen:2023szb}. Thus, $\hat{P}_\chi$ plays the role of a \qmarks{reduction map} from $\mathcal{F}$ and $\mathcal{F}_\chi$. However, note that the reduction map $\hat{P}_\chi$ is a non-invertible projector, which explicitly highlights the distinct physical content of $\mathcal{F}$ and $\mathcal{F}_\chi$. In other words, $\mathcal{F}_\chi$ does not merely describe $\mathcal{F}$ from the perspective of $\chi$; rather, it is an entirely distinct space with different physical properties, and in particular with different commutation relations, without any requirement of compatibility with $\mathcal{F}$. In contrast, note that in the QRF context the perspective-neutral and reduced descriptions are in fact equivalent, and the reduction map is invertible, since it is a gauge redundancy that has been removed in the reduced description \cite{delaHamette:2021oex}. In this sense, the terminology “perspective-reduced” and “reduction maps” introduced here should be interpreted with appropriate caution, keeping in mind the important differences between the GFT and the QRF frameworks.

Since $\hat{P}_\chi$ is a projector, it can be used to define an effect density on $\mathcal{F}_\chi$ through
\begin{equation}
    \hat{\tilde{E}}^{(P)}_\chi\equiv \diff\chi\hat{P}_\chi\,.
\end{equation}
This is clearly positive and additive, and it also satisfies
\begin{align}    
    \int \hat{\tilde{E}}^{(P)}_\chi \hat{\tilde{\varphi}}_{\vec{\kappa}}(\chi)
    &= \sqrt{\delta(0)} \int\diff\chi \hat{P}_\chi \hat{P}_\chi \hat{\varphi}_{\vec{\kappa}}(\chi)\hat{P}_\chi
    = \sqrt{\delta(0)} \int\diff\chi \hat{P}_\chi \hat{{\varphi}}_{\vec{\kappa}}(\chi)\hat{P}_\chi\hat{P}_\chi
    \nonumber\\
    &= \int \hat{\tilde{\varphi}}_{\vec{\kappa}}(\chi)\hat{\tilde{E}}^{(P)}_\chi\,,
\end{align}
and similarly for $\hat{\tilde{\varphi}}^\dagger$. Hence, $\hat{\tilde{E}}^{(P)}_\chi$ generates a POVM on $\mathcal{F}_\chi$. 

We can therefore use $\hat{\tilde{E}}^{(P)}_\chi$ (or, equivalently, $\hat{P}_\chi$) to define perspective-reduced observables on $\mathcal{F}_\chi$. Given an $(n,m)$-body observable $\hat{O}^{n,m}[\hat{\varphi},\hat{\varphi}^\dagger]$ as defined in \eqref{eqn:nmbodyobs}, one can construct its perspective-reduced counterpart on $\mathcal{F}_\chi$ by projection:
\begin{equation}\label{eqn:projectedop}
    \hat{O}^{(n,m)}_{\chi,P}\equiv \hat P_\chi \, \hat{O}^{(n,m)}[\hat{\varphi},\hat{\varphi}^\dagger] \, \hat P_\chi\,.
\end{equation}
Note that, up to an overall (distributional) factor, this is equivalent to defining the perspective-reduced observables as sharply localized operators by replacing the field operators by the projected field operators \eqref{eqn:projectedfieldops}:
\begin{equation}\label{eqn:projectedoperators2}
    \hat{\tilde{O}}^{(n,m)}_\chi\equiv \int \left[ \prod_{i=1}^{n} \diff\vec{g}_i\, \hat{\tilde{\varphi}}^\dagger(\vec{g}_i,\chi) \right] \left[ \prod_{j=1}^m \diff\vec{g}^{{}\,\prime}_j \, \hat{\tilde{\varphi}}(\vec{g}^{{}\,\prime}_j,\chi)\right] O^{(n,m)}(\chi)\,,
\end{equation}
where $O^{(n,m)}(\chi)$ can depend on all $\vec{g}_i$, $\vec{g}^{{}\,\prime}_j$, but all $n+m$ arguments for $\chi_i$ are set to $\chi$. The relation between these two definitions, as shown in \ref{app:proj}, is
\begin{equation}\label{eqn:projectedequivalence}
    \hat{\tilde{O}}^{(n,m)}_\chi = [\delta(0)]^{-\frac{n+m}{2}} \, \hat{O}^{(n,m)}_{\chi,P} \,.
\end{equation}
For example, for the scalar field and its momentum (with a factor ordering so it is hermitian)
\begin{equation}
    \hat{\tilde{X}}_\chi=\chi \sum_{\vec{\kappa}}\hat{\tilde{\varphi}}^\dagger_{\vec{\kappa}}(\chi)\hat{\tilde{\varphi}}_{\vec{\kappa}}(\chi)\,,\qquad \hat{\tilde{\Pi}}_\chi=-\frac{i}{2}\sum_{\vec{\kappa}}\left\{\hat{\tilde{\varphi}}^\dagger_{\vec{\kappa}}(\chi)\partial_\chi\hat{\tilde{\varphi}}_{\vec{\kappa}}(\chi)-\partial_\chi[\hat{\tilde{\varphi}}^\dagger_{\vec{\kappa}}(\chi)]\hat{\tilde{\varphi}}_{\vec{\kappa}}(\chi)\right\},
\end{equation}
while the action of $\hat{\tilde{X}}_\chi$ on an $n$-body state of $\mathcal{F}_\chi$ is
\begin{equation}
    \hat{\tilde{X}}_\chi\ket{\vec{\kappa}_1,\chi,\dots, \vec{\kappa}_n,\chi}=n\chi \ket{\vec{\kappa}_1,\chi,\dots, \vec{\kappa}_n,\chi}.
\end{equation}
On the other hand, the action of the first moment of $\hat{\tilde{E}}^{(P)}_\chi(\diff\chi)$, i.e., $\hat{\tilde{\chi}}_\chi(\diff\chi)\equiv\diff\chi\, \chi \hat{P}_\chi$ on the same state is
\begin{equation}
    \hat{\tilde{\chi}}_\chi(\diff\chi)\ket{\vec{\kappa}_1,\chi,\dots, \vec{\kappa}_n,\chi}=\diff\chi\,\chi \ket{\vec{\kappa}_1,\chi,\dots, \vec{\kappa}_n,\chi},
\end{equation}
showing that $\hat{\tilde{\chi}}_\chi(\diff\chi)$ can indeed be identified with the intensive scalar field operator (density) on $\mathcal{F}_\chi$. 

Note, however, that the $\hat{\tilde{E}}^{(P)}_\chi$ is \textit{not} covariant with respect to $\hat{\tilde{\Pi}}_\chi$. To see this, consider
\begin{align*}
    \left[\hat{\tilde{E}}^{(P)}_\chi,\hat{\tilde{\Pi}}_\chi\right]=\diff\chi\left[\hat{P}_\chi,\hat{\tilde{\Pi}}_\chi\right]&=\frac{-i}{2\delta(0)}\diff\chi\left[\hat{P}_\chi,\sum_{\vec{\kappa}}\hat{P}_\chi\hat{\varphi}^\dagger_{\vec{\kappa}}(\chi)\hat{P}_\chi\partial_\chi(\hat{P}_\chi\hat{\varphi}_{\vec{\kappa}}(\chi)\hat{P}_\chi)-\text{h.c.}\right];
\end{align*}
the first term gives
\begin{align*}
    &-i\diff\chi\left\{\hat{\tilde{\Pi}}_\chi-\frac{1}{\delta(0)} \sum_{\vec{\kappa}}\hat{P}_\chi\hat{\varphi}^\dagger_{\vec{\kappa}}(\chi)\hat{P}_\chi\partial_\chi(\hat{P}_\chi\hat{\varphi}_{\vec{\kappa}}(\chi)\hat{P}_\chi)\hat{P}_\chi\right\} \\
    &\quad\quad= -i \diff\chi \left\{ \hat{\tilde{\Pi}}_\chi - \hat{\tilde{\Pi}}_\chi +\hat{\tilde{N}}_\chi \partial_\chi \hat{P}_\chi \right\}
    = -i \hat{\tilde{N}}_\chi\partial_\chi\hat{\tilde{E}}_\chi\,,
\end{align*}
and because $\hat{\tilde{E}}^{(P)}_\chi$ is hermitian, $[\hat{\tilde{E}}^{(P)}_\chi,\hat{A}^\dagger]=-[\hat{\tilde{E}}^{(P)}_\chi,\hat{A}]^\dagger$ for all $\hat A$, so
\begin{equation}
    \left[\hat{\tilde{E}}^{(P)}_\chi,\hat{\tilde{\Pi}}_\chi\right]=-\frac{i}{2}\left[\hat{\tilde{N}}_\chi\partial_\chi\hat{\tilde{E}}^{(P)}_\chi+(\partial_\chi\hat{\tilde{E}}^{(P)}_\chi)\hat{\tilde{N}}_\chi\right]=-i\hat{\tilde{N}}_\chi\partial_\chi\hat{\tilde{E}}^{(P)}_\chi,
\end{equation}
since $\hat{\tilde{N}}_\chi$ and $\partial_\chi\hat{\tilde{E}}^{(P)}_\chi$ commute. Thus, as stated above, $\hat{\tilde{E}}^{(P)}_\chi$ is not covariant with respect to $\hat{\tilde{\Pi}}_\chi$.

On the other hand, defining
\begin{equation}
    \hat{\pi}_\chi\equiv -i\left[\hat{P}_\chi\partial_\chi\hat{P}_\chi-(\partial_\chi\hat{P}_\chi)\hat{P}_\chi\right],
\end{equation}
it then follows that
\begin{equation}
    \left[\hat{\tilde{E}}^{(P)}_\chi,\hat{\pi}_\chi\right]=-i\partial_\chi\hat{\tilde{E}}^{(P)}_\chi\,,
\end{equation}
so $\hat{\tilde{E}}^{(P)}_\chi$ is covariant with respect to $\tilde{\hat{\pi}}_\chi$. Note, however, that the matrix elements of $\hat{\pi}_\chi$ are identically zero on $\mathcal{F}_\chi$; indeed, its projected version $\hat{\tilde{\pi}}_\chi=P_\chi\hat{\pi}_\chi P_\chi$ vanishes identically on $\mathcal{F}_\chi$. If the synchronized space $\mathcal{F}_\chi$ is interpreted as a \qmarks{reduced} sector of the GFT Fock space, it is then natural to interpret $\hat{\pi}_\chi$ as part of the clock degrees of freedom, which therefore have trivial action on the reduced space (and these degrees of freedom are not accessible from the reduced space $\mathcal{F}_\chi$). Note, however, that this picture can only be qualitatively true, as there is no \qmarks{time reparametrization constraint} in the theory to be solved (although see \cite{Calcinari:2024pek} for a proposal for such a constraint in GFT). 

Finally, let us consider the dynamics in the synchronized sector $\mathcal{F}_\chi$. The dynamics can be derived either from a synchronized Lagrangian or a synchronized Hamiltonian, related by a Legendre transform. From the Hamiltonian perspective, for concreteness, consider the simple Hamiltonian (similar results hold for more general systems)
\begin{equation}
    \hat{\tilde{\mathcal{H}}}_\chi\equiv \frac{1}{2}\sum_{\vec{\kappa}}\hat{\tilde{\Pi}}_{\Phi,\vec{\kappa}}(\chi)\hat{\tilde{\Pi}}_{\Phi,\vec{\kappa}}(\chi)+U[\hat{\tilde{\Phi}}]\,,
\end{equation} 
where the field and momentum operators are defined in terms of $\hat{\tilde{\varphi}}$ and $\hat{\tilde{\varphi}}^\dag$ as
\begin{equation}\label{eqn:reducedoperators}
    \hat{\tilde{\Phi}}_{\vec{\kappa}}(\chi)=\frac{1}{\sqrt{2}}\left(\hat{\tilde{\varphi}}_{\vec{\kappa}}(\chi)+\hat{\tilde{\varphi}}^\dagger_{\vec{\kappa}}(\chi)\right),\qquad  \hat{\tilde{\Pi}}_{\Phi,\vec{\kappa}}(\chi)=\frac{-i}{\sqrt{2}}\left(\hat{\tilde{\varphi}}_{\vec{\kappa}}(\chi)-\hat{\tilde{\varphi}}^\dagger_{\vec{\kappa}}(\chi)\right),
\end{equation}
and the corresponding GFT field $\tilde\Phi$ is taken to be real. Recall, as discussed below \eqref{eqn:commutationrelationsk}, that there exist different proposals for the relation between the GFT field operators with the creation and annihilation operators. In the case of a real GFT field $\Phi$ whose field operator is given by $\hat \Phi_{\vec{\kappa}}(\chi) = (1/\sqrt{2}) [\hat \varphi_{\vec{\kappa}}(\chi) + \hat\varphi^\dagger_{\vec{\kappa}}(\chi)]$, then the first equation in \eqref{eqn:reducedoperators} is nothing but the \qmarks{perspective-reduced} version of the \qmarks{perspective-neutral} field operator in the spin representation. On the other hand, if the GFT field was originally complex, then in the Hamiltonian constructed to provide the dynamics for the synchronized system, the original complex GFT field is reduced to a synchronized real GFT field $\tilde\Phi$.

In either case, defining the Lagrangian to satisfy
\begin{equation}
    \hat{\tilde{\Pi}}_{\Phi,\vec{\kappa}}(\chi)\equiv \frac{\partial \hat{\tilde{\mathcal{L}}}_\chi}{\partial\Big(\partial_\chi \hat{\tilde{\Phi}}_{\vec{\kappa}}(\chi) \Big)} \equiv\partial_\chi \hat{\tilde{\Phi}}_{\vec{\kappa}}(\chi)\,,
\end{equation}
the Legendre transform of the Hamiltonian gives
\begin{equation}\label{eqn:projectedlagrangian}
    \hat{\tilde{\mathcal{L}}}_\chi=\frac{1}{2}\sum_{\vec{\kappa}}\left(\partial_\chi \hat{\tilde{\Phi}}_{\vec{\kappa}}(\chi) \right)^2-U[\hat{\tilde{\Phi}}]\,,
\end{equation}
matching the Lagrangian typically considered in perspective-reduced approaches to GFT \cite{Wilson-Ewing:2018mrp, Gielen:2019nyq, Gielen:2019kae, Gielen:2020fgi, Gielen:2023szb}.

\section{Discussion and conclusions}
\label{sec:conclusions}

In this work, following recent advances in the study of dynamical and quantum reference frames \cite{Hoehn:2019fsy, Hoehn:2020epv, Loveridge:2017pcv, Vanrietvelde:2018dit, Vanrietvelde:2018pgb, Hoehn:2021flk, delaHamette:2021oex, Carette:2023wpz, Hoehn:2023ehz}, we constructed relational observables that act on the perspective-neutral GFT Fock space by conditioning (on-shell) \qmarks{partial} observables on frame orientations. These orientations are encoded in covariant positive operator-valued measures (POVMs), which are conceptually similar to the event POVMs used to quantize spacetime coordinates in QFT \cite{Toller:1997pc, Toller:1998wf, Toller:1998vk, Giannitrapani:1998pm}. Since the GFT field domain represents a space of field values rather than spacetime itself, event POVMs in the GFT context provide information about the quantum fields within the system under consideration. This was demonstrated by explicitly constructing POVMs associated with minimally-coupled massless scalar fields and showing that the first moment of the corresponding effect density offers a consistent definition of an intensive observable for the scalar field. Through the on-shell conditioning process, it is possible to explicitly construct relational observables with respect to a massless scalar field quantum reference frame.

These observables represent a significant improvement over previous definitions of relational observables in GFT \cite{Oriti:2016qtz, Gielen:2016dss, Gielen:2017eco, Gielen:2018xph, Gielen:2018fqv, Gerhardt:2018byq, Wilson-Ewing:2018mrp, Gielen:2019nyq, Gielen:2019kae, Gielen:2020fgi, Gielen:2023szb, Marchetti:2020umh, Marchetti:2020qsq, Gielen:2021vdd, Marchetti:2021gcv, Jercher:2021bie, Jercher:2023kfr, Jercher:2023nxa}, as the quantum properties of the scalar field reference frame are fully incorporated into the relational observables through the appropriate POVMs. Additionally, these observables are derived from a quantum perspective-neutral description of the GFT, without requiring any prior perspective reduction at the level of the classical theory. In fact, we have demonstrated that such a perspective-reduced formalism \cite{Wilson-Ewing:2018mrp, Gielen:2019nyq, Gielen:2019kae, Gielen:2020fgi, Gielen:2023szb} can be recovered within this approach by restricting to a specific, \qmarks{ultra-localized} class of POVMs associated with the QRF of interest. Furthermore, we have shown that the intrinsic uncertainties in the measurement process, captured by the POVMs, when averaged over suitable states give the state-dependent effective relational observables defined in~\cite{Marchetti:2020umh, Marchetti:2020qsq, Marchetti:2021gcv, Jercher:2021bie, Jercher:2023kfr, Jercher:2023nxa}.

Importantly, these conceptual and technical advancements in defining relational observables through POVMs incur virtually no additional cost at the level of expectation values over coherent states in the continuum (and classical) limit, as these expectation values match those obtained using previous definitions of relational observables available in the literature. This is particularly significant as it ensures that the relational observables developed here can be effectively employed to extract continuum physics, especially in a cosmological context, without any modification to the main results of previous work. Nonetheless, these relational observables generally differ significantly from previous definitions, with the differences becoming more pronounced away from the semi-classical, continuum regime. Thus, it would be valuable to explore these differences in a more systematic manner. We leave this investigation for future work.

Additionally, the systematic method presented here for turning operators in the GFT Fock space into relational observables may provide access to new relational observables of physical interest (the prior absence of such observables had hindered phenomenological progress, particularly in the cosmological context \cite{Jercher:2023nxa,Jercher:2023kfr}).

Furthermore, it is important to emphasize that the procedure for defining relational observables within GFT developed here can, in principle, be applied to any type of QRF, not just minimally-coupled massless scalar field frames. Although this may be technically more complex, having access to QRFs (represented by appropriate POVMs) with different physical properties would enable a concrete study of how to switch between different QRFs. This, in turn, would be essential for investigating the fundamental question of QRF covariance within the context of full QG.

Lastly, we note that although the procedure for defining relational observables in GFT described here can be naturally implemented within a Fock quantization, it is not confined to this framework. Since different formulations of GFT are closely connected to different approaches to QG (as discussed in Sec.~\ref{sec:gfts}), this opens the possibility of applying the procedures and techniques developed here to other frameworks, such as canonical LQG and spinfoam models.

\appendix

\section{Explicit computations}
\label{app:computations}

\subsection{Essentially self-adjoint}
\label{app:sa}

In this appendix we prove that the operator $\hat\chi$ given in \eqref{eqn:scalarfieldoperator} is essentially self-adjoint.

From \eqref{eqn:scalarfieldoperatoraction}, it is clear that the $n$-particle states are eigenvectors of $\hat{\chi}$ with real eigenvalues $\lambda_\chi$. As a consequence, the range of $(\hat{\chi}\pm i)$ is dense. To see this, suppose that there exists a vector $\ket{\psi}$ orthogonal to the range of $(\hat{\chi}\pm i)$, i.e., for any $\ket{\phi}\in\mathcal{F}$, $\bra{\psi} (\hat{\chi}\pm i) \ket{\phi}=0$. Now take $\ket{\phi}$ to be an $n$-particle state, $\ket{n}$ (using a simplified notation here that suppresses the explicit dependence of the $n$-particle state on all the quantum numbers). Then the above condition becomes $(\lambda_\chi\pm i)\braket{\psi}{ n}=0$, for all $n$-particle states $\ket{n}$ (including the vacuum state $\ket{0}$ for which $\lambda_\chi=0$). But since $\lambda_\chi\in\mathbb{R}$ and the $n$-particle states form a basis for $\mathcal{F}$, we conclude that the above condition implies $\ket{\psi}=0$. Thus, the range of $(\hat{\chi}\pm i)$ is dense, and it follows that the operator $\hat\chi$ is essentially self-adjoint: it has a unique self-adjoint extension. 

\subsection{Covariance}
\label{app:cov}

In this appendix we prove equation \eqref{eqn:commutatormomentum}. Since
\begin{align}\label{eqn:commutatorproof}
    \left[\hat{E}_{\tilde \chi}(\diff\tilde \chi),\hat{\Pi}\right]&=i\diff\tilde\chi\sum_{n=1}^\infty\sum_{\vec{\kappa},\vec{\kappa}_1,\dots\vec{\kappa}_n}\int\diff\chi\left[\prod_{j=1}^n\diff\chi_j\right]\frac{1}{n} \sum_{k=1}^n\delta(\chi_k-\tilde \chi)\nonumber\\
    &\quad\times\left[\hat{\varphi}_{\vec{\kappa}}^\dagger(\chi)\partial_\chi\hat{\varphi}_{\vec{\kappa}}(\chi),F^{(n)}_{\vec{\kappa}_1,\dots,\vec{\kappa}_n}(\chi_1,\dots,\chi_n)\right],
\end{align}
the proof boils down to computing the following commutators: 
\begin{align}\label{eqn:commutatorappendix}
 \!\!   \left[\hat{\varphi}_{\vec{\kappa}}^\dagger(\chi)\partial_\chi\hat{\varphi}_{\vec{\kappa}}(\chi),F^{(n)}_{\vec{\kappa}_1,\dots,\vec{\kappa}_n}(\chi_1,\dots,\chi_n)\right]&= \, \hat{\varphi}_{\vec{\kappa}}^\dagger(\chi)\left[\partial_\chi\hat{\varphi}_{\vec{\kappa}}(\chi),F^{(n)}_{\vec{\kappa}_1,\dots,\vec{\kappa}_n}(\chi_1,\dots,\chi_n)\right]\nonumber\\
    &\quad+\left[\hat{\varphi}_{\vec{\kappa}}^\dagger(\chi),F^{(n)}_{\vec{\kappa}_1,\dots,\vec{\kappa}_n}(\chi_1,\dots,\chi_n)\right]\partial_\chi\hat{\varphi}_{\vec{\kappa}}(\chi)\,.
\end{align}
To this purpose, the following identity is useful:
\begin{align}\label{eqn:commutatorfa}
   \left[\hat{F}^{(n)}_{\vec{\kappa}_1,\dots,\vec{\kappa}_n}(\chi_1,\dots,\chi_n),\hat{\varphi}_{\vec{\kappa}}(\chi)\right]&=-\frac{1}{n}\sum_{i=1}^n\delta_{\vec{\kappa}_i,\vec{\kappa}}\delta(\chi_i-\chi)\hat{F}^{n-1}_{[j\neq i]}\hat{\varphi}_{\vec{\kappa}_i}(\chi_i)\nonumber\\
   &\qquad + \hat{F}^{(n)}_{\vec{\kappa}_1,\dots,\vec{\kappa}_n}(\chi_1,\dots,\chi_n)\hat{\varphi}_{\vec{\kappa}}(\chi)\,,
\end{align}
where $\hat{F}^{(n-1)}_{[j\neq i]}$ is shorthand notation for $\hat{F}^{(n-1)}_{\vec{\kappa}_1\dots,\vec{\kappa}_{i-1},\vec{\kappa}_{i+1},\dots, \vec{\kappa}_n}(\chi_1,\dots,\chi_{i-1},\chi_{i+1},\dots,\chi_n)$. From this and $\left[\hat{F}^{(n)}_{\vec{\kappa}_1,\dots,\vec{\kappa}_n}(\chi_1,\dots,\chi_n),\hat{\varphi}^\dagger_{\vec{\kappa}}(\chi)\right] = -\left[\hat{F}^{(n)}_{\vec{\kappa}_1,\dots,\vec{\kappa}_n}(\chi_1,\dots,\chi_n),\hat{\varphi}_{\vec{\kappa}}(\chi)\right]^\dagger$, a direct calculation shows that \eqref{eqn:commutatorappendix} is given by
\begin{align}\label{eqn:commutatorappendix1}
\!\!    \left[\hat{\varphi}_{\vec{\kappa}}^\dagger(\chi)\partial_\chi\hat{\varphi}_{\vec{\kappa}}(\chi),F^{(n)}_{\vec{\kappa}_1,\dots,\vec{\kappa}_n}(\chi_1,\dots,\chi_n)\right]&=
\frac{1}{n} \sum_{i=1}^n \delta_{\vec{\kappa}_i,\vec{\kappa}} \bigg( \Big(\partial_\chi\delta(\chi_i-\chi) \Big) \hat{\varphi}_{\vec{\kappa}}^\dagger(\chi) \hat{F}^{n-1}_{[j\neq i]} \hat{\varphi}_{\vec{\kappa}_i}(\chi_i) \nonumber\\
    &\qquad \quad
    - \delta(\chi-\chi_i)\hat{\varphi}^\dagger_{\vec{\kappa}_i}(\chi_i)\hat{F}^{n-1}_{[j\neq i]}\partial_\chi\hat{\varphi}_{\vec{\kappa}}(\chi) \bigg).
\end{align}
Substituting this in \eqref{eqn:commutatorproof}, integrating the first term by parts, and then integrating over $\chi$ to remove the delta functions that arose in \eqref{eqn:commutatorappendix1},
\begin{equation}
\!\!\!    \left[\hat{E}_{\tilde \chi}(\diff\tilde\chi),\hat{\Pi}\right] = -i\diff\tilde\chi\sum_{n=1}^\infty\sum_{\vec{\kappa}_1,\dots\vec{\kappa}_n} \! \int \! \left[\prod_{j=1}^n\diff\chi_j\right] \! \frac{1}{n} \! \sum_{k,\ell=1} ^n \! \delta(\chi_k-\tilde\chi) \, \partial_{\chi_\ell} F^{(n)}_{\! \vec{\kappa}_1,\dots,\vec{\kappa}_n} \! (\chi_1,\dots,\chi_n)\,.
\end{equation}
In the case that $k \neq \ell$, the quantity $\partial_{\chi_\ell} F^{(n)}_{\vec \kappa_i}(\chi_i)$ integrated over $\chi_\ell$ provides only boundary contributions, which are imposed to vanish. The remaining terms with $k=\ell$ give \eqref{eqn:commutatormomentum}.

\subsection{Expectation value on coherent states}
\label{app:coherent}

In this appendix, we derive equation \eqref{eqn:expvaluegenericcoherent2}. We start from equation \eqref{eqn:expvaluecoherent1}, which, by using the definitions \eqref{eqn:notation1}, gives
\begin{align}\label{eqn:expvalueseriesform}
    \langle\hat{O}_\chi(\diff\chi)\rangle_\sigma&=\vert\mathcal{N}_\sigma\vert^2\diff\chi\left\{O_\sigma(\chi)\sum_{n=1}^\infty\frac{1}{n!}N_\sigma^{n-1}+\langle\hat{O}\rangle_\sigma N_\sigma(\chi)\sum_{n=2}^\infty\frac{n-1}{n!}N_\sigma^{n-2}\right\}\nonumber\\
    &=\vert\mathcal{N}_\sigma\vert^2\diff\chi\left\{O_\sigma(\chi)\sum_{n=0}^\infty\frac{1}{(n+1)!}N_\sigma^n+\langle\hat{O}\rangle_\sigma N_\sigma(\chi)\sum_{n=0}^\infty\frac{n+1}{(n+2)!}N_\sigma^{n}\right\}.
\end{align}
This can be further simplified using $\vert\mathcal{N}_\sigma\vert^2=e^{-N_\sigma}=\sum_{n=0}^\infty (-1)^n N_\sigma^n/n!$, and that the product of two convergent series is
\begin{equation}
    \left(\sum_{n=0}^\infty a_nx^n\right)\left(\sum_{n=0}^\infty b_nx^n\right)=\sum_{n=0}^\infty c_n x^n\,,\qquad c_n=\sum_{k=0}^n a_kb_{n-k}\,.
\end{equation}
In this case, the required coefficients $c_n$ are given by the sums
\begin{align}
    c_n^{(1)} =& \, \sum_{k=0}^n \frac{1}{(k+1)!} \cdot \frac{(-1)^{n-k}}{(n-k)!}
    = \frac{(-1)^n}{(n+1)!} \sum_{k=0}^n (-1)^k \binom{n+1}{k+1} \,, \\
    c_n^{(2)} =& \, \sum_{k=0}^n \frac{k+1}{(k+2)!} \frac{(-1)^{n-k}}{(n-k)!}
    =\frac{(-1)^n}{(n+2)!} \sum_{k=0}^n (-1)^k (k+1) \binom{n+2}{k+2} \,.
\end{align}
These weighted sums of binomial coefficients are easily evaluated using the identity
\begin{equation}
\binom{n+1}{k+1} = \binom{n}{k} + \binom{n}{k+1},
\end{equation}
with the result that $c_n^{(1)} = (-1)^n/(n+1)!$ and $c_n^{(2)} = (-1)^n/(n+2)!$, from which
\begin{subequations}\label{eqn:series}
\begin{align}
    \!\!
    \vert\mathcal{N}_\sigma\vert^2\sum_{n=0}^\infty\frac{1}{(n+1)!}N_\sigma^n &=
    \! \sum_{n=0}^\infty \frac{(-1)^n}{(n+1)!} N_\sigma^n =
    -\frac{1}{N_\sigma}\sum_{n=1}^\infty\frac{(-1)^n}{n!}N_\sigma^n
    =-\frac{\vert\mathcal{N}_\sigma\vert^2-1}{N_\sigma}\,,\\
    \!\!\!\!
    \vert\mathcal{N}_\sigma\vert^2\sum_{n=0}^\infty\frac{n+1}{(n+2)!}N_\sigma^n &=\! \sum_{n=0}^\infty \frac{(-1)^n}{(n+2)!} N_\sigma^n =
    \frac{1}{N_\sigma^2}\sum_{n=2}^\infty\frac{(-1)^n}{n!}N_\sigma^n=\frac{\vert\mathcal{N}_\sigma\vert^2 + N_\sigma \! - 1}{N_\sigma^2}.
\end{align}
\end{subequations}
Substituting these relations back in \eqref{eqn:expvalueseriesform} gives \eqref{eqn:expvaluegenericcoherent2}.

\subsection{Equivalence between projected and sharply localized observables}
\label{app:proj}

In this appendix, we prove equation \eqref{eqn:projectedequivalence}. To show this, first note that
\begin{subequations}
\label{eqn:pphitilde}
\begin{align}
    \hat{\varphi}_{\vec{\kappa}}(\chi')\hat{P}_\chi
    &= \hat{\varphi}_{\vec{\kappa}}(\chi')\hat{P}_\chi\hat{P}_\chi \nonumber \\
    &=\frac{\delta(\chi-\chi')}{\delta(0)}\sum_{n=1}^\infty\sum_{\vec{\kappa}_1,\dots,\vec{\kappa}_{n-1}}[\delta(0)]^{-(n-1)}\hat{F}^{(n-1)}_{\vec{\kappa}_1,\dots,\vec{\kappa}_{n-1}}(\chi)\hat{\varphi}_{\vec{\kappa}}(\chi)\hat{P}_\chi\nonumber\\
    &=\frac{\delta(\chi-\chi')}{\delta(0)}\sum_{n=0}^\infty\sum_{\vec{\kappa}_1,\dots,\vec{\kappa}_{n}}[\delta(0)]^{-n}\hat{F}^{(n)}_{\vec{\kappa}_1,\dots,\vec{\kappa}_{n}}(\chi)\hat{\varphi}_{\vec{\kappa}}(\chi)\hat{P}_\chi \nonumber\\
    &=\frac{\delta(\chi-\chi')}{\delta(0)}\hat{P}_\chi\hat{\varphi}_{\vec{\kappa}}(\chi)\hat{P}_\chi\nonumber\\
    &=\frac{\delta(\chi-\chi')}{\sqrt{\delta(0)}}\hat{\tilde{\varphi}}_{\vec{\kappa}}(\chi) % \nonumber \\ &
    =\frac{\delta(\chi-\chi')}{\sqrt{\delta(0)}}\hat{P}_\chi\hat{\tilde{\varphi}}_{\vec{\kappa}}(\chi)\,,
\end{align}
where $F^{(0)}\equiv \ket{0}\bra{0}$ and where the bosonic statistics of the GFT field operator are used in the third line. Also, taking the hermitian conjugate of this equation gives
\begin{equation}
    \hat{P}_\chi\hat{\varphi}_{\vec{\kappa}}^\dagger(\chi')=\frac{\delta(\chi-\chi')}{\sqrt{\delta(0)}}\hat{\tilde{\varphi}}^\dagger_{\vec{\kappa}}(\chi)=\frac{\delta(\chi-\chi')}{\sqrt{\delta(0)}}\hat{\tilde{\varphi}}_{\vec{\kappa}}(\chi)\hat{P}_\chi\,.
\end{equation}
\end{subequations}
The two equations \eqref{eqn:pphitilde} can be used iteratively in \eqref{eqn:projectedop}. Each integral in $\hat O$ over scalar field arguments is localized by the delta functions in the numerator of the above equations, and each iteration produces $[\delta(0)]^{-1/2}$; the result is
\begin{equation}
    \hat{O}^{(n,m)}_{\chi,P}=[\delta(0)]^{-\frac{n+m}{2}} \, \hat{\tilde{O}}^{(n,m)}_{\chi}\,.
\end{equation}

\bigskip

\noindent
\textit{Acknowledgments:}~
The authors would like to thank Branimir \'Ca\'ci\'c, Fabio Mele, Andreas Pithis and in particular Philipp H\"ohn for helpful discussions and comments.
This work was supported in part by the Natural Sciences and Engineering Research Council of Canada.
L.M.~also acknowledges support from the Atlantic Association for Research in Mathematical Sciences, from the Okinawa Institute of Science and Technology Graduate University, and from the John Templeton Foundation, through ID\# 62312 grant as part of the \href{https://www.templeton.org/grant/the-quantum-information-structure-of-spacetime-qiss-second-phase}{\textit{`The Quantum Information Structure of Spacetime'} Project (QISS)}.

\section*{References}

\raggedright

\end{document}